\documentclass[doublespacing, 12 point]{article}
\usepackage[left=1in, right=1in, top=1in, bottom=1in]{geometry}
\usepackage{amsfonts, amsmath, amssymb}
\usepackage{upgreek}
\usepackage{lineno}
\usepackage{graphicx}
\usepackage[usenames, dvipsnames]{color}
\usepackage[allcolors=blue]{hyperref}
\hypersetup{colorlinks,bookmarksnumbered}

\usepackage[normalem]{ulem}
\usepackage{setspace}
\usepackage{enumerate}
\usepackage{hyphenat}
\usepackage{array}
\usepackage{nicefrac}

\usepackage{multirow}
\usepackage{longtable}
\usepackage{booktabs}
\usepackage{makecell}

\newcolumntype{M}[1]{>{\centering\arraybackslash}m{#1}}

\usepackage{lipsum}

\newcommand{\up}{\textsuperscript}

\newcommand{\pr}{$P_{R}$ }
\newcommand{\pfr}{$P_{FR}$ }

\begin{document}
\title{Suppression of the Spectral Cross Talk of Optogenetic Switching by Stimulated Depletion Quenching. Theoretical Analysis. }%\label{ch:OG}
\author{Zachary Quine, Alexei Goun, Herschel Rabitz}
\date{}

%\author{Zachary Quine}
%\email{zquine@princeton.edu}
%\affiliation{Department of Chemistry, Princeton University, Princeton, NJ 08544, USA}
%
%\author{Alexei Goun}
%\email{agoun@princeton.edu}
%\affiliation{Department of Chemistry, Princeton University, Princeton, NJ 08544, USA}
%
%\author{Herschel A. Rabitz}
%\email{hrabitz@princeton.edu}
%\affiliation{Department of Chemistry, Princeton University, Princeton, NJ 08544, USA}

\maketitle

\begin{abstract}

Optogenetics is a rapidly growing field of biotechnology, potentially allowing a deeper understanding and control of complex biological networks. The major challenge is the multiplexed control of several optogenetic components in the presence of significant spectral cross talk. We propose and demonstrate through simulations a new control approach of Stimulated Depletion Quenching. This approach is applied to the phytochrome Cph8 bidirectional optogenetic switch, and the results show significant improvement of its dynamic range.

\end{abstract}

\section*{Introduction and Background}\label{Sec:OG:IntroBG}

\space Optical control and detection are of very high utility in biological sciences. Through ongoing development of optogenetics it is becoming possible to control and interrogate biological behavior in space and time. There are multiple families of light sensitive proteins that can be connected to a variety of effectors thus providing optogenetic switches controlling a broad range of biological functions \cite{Deisseroth2011,Pastrana2011,Fenno2011}. Among the widely used families of proteins there are light-gated ion channels, represented by the Channelrhodopsin family of proteins \cite{zhang2011microbial,hegemann2013channelrhodopsins}, LOV domains \cite{Pudasaini2015} that can serve as modulators of gene activity, and the near-IR sensitive Phytochrome family \cite{Levskaya2005,Burgie2014Rev}. The growing set of light sensitive optogenetic switches, together with a palette of fluorescent protein reporters opens up an opportunity for advanced control and understanding of complex biological networks from complex disease states to biologically inspired production of organic molecules. \par 
 One of the most pressing issues is the ability to simultaneously control multiple optogenetic switches with a high degree of fidelity. The standard approach to multiplexed optical access is the excitation of the optogenetic protein by a carefully chosen narrowband light source that minimizes the absoprtion and photoexcitation of often competing optogenetic channels. In this approach, the temporal structure of light plays a limited role. As long as the light pulse duration is shorter than relevant biological time scale, the only relevant parameter is the integrated exposure. We intend to explore the regime of light-matter interaction where the characteristic pulse duration is comparable or shorter to the time scale of the intramolecular switching processes. The application of ultrashort laser pulses, picosecond and shorter, has allowed enhanced control in a wide range of physical and chemical processes including two photon absorption, Raman excitation of molecular vibrations, the study of light harvesting complexes, etc. In this paper we evaluate the opportunity to improve the operation of optogenetic switches by ultrafast optical control. \par
The most flexible route to explore the complex ultrafast pulse control capabilities is through the use of optical pulse shaping. In such an approach an entire space of optical pulse shapes of a given spectral bandwidth and spectral resolution can be utilized. Despite the very high dimensionality of the search space, the expected favorable topological properties of the control can allow for creation of efficient search algorithms. In the laboratory a suitable algorithm can guide the variation of the applied pulse shape followed by observation of the change of a targeted experimental characteristic such as the efficiency of optogenetic switching or the fidelity of control over multiple switches. At the same time the application of the such control algorithms requires a high fidelity feedback signal to update "shape" of the control light pulse ideally leading to improved switching performance. At present, a high-fidelity readout of the state of an optogenetic protein is challenging. Thus, we will consider an optical control approach that we refer as “stimulated depletion quenching” (SDQ) with a simpler temporal structure. In SDQ approach we can systematically explore the entire control space. We will utilize the near-IR sensitive Phytochrome Cph8 optogenetic protein system as a basis of our theoretical model. A model based on a kinetic mechanism will be utilized, as the parameters of this system are well characterized, and its spectral characteristics are convenient for the experimental validation of the control approach \cite{2018arXiv181011432Q}. \par

    The theory and simulations described here present a feasibility study of this proposed SDQ mechanism, 
This theoretical work forms a first step toward the ultimate goals of achieving maximal dynamic range control of a single optogenetic switch and simultaneous independent control of multiple optogenetic switches in living cells by multiplexed, optimal non-linear photoswitching. The remainder of this introduction section contains additional background information on the Cph8 phytochrome based switch, the photoisomerization of the phycocyanobilin (PCB) chromophore associated with Cph8 , and a more detailed description of the proposed SDQ technique.
 Then, we will describe a model for simulations assessing the feasibility of an appropriate experiment and exploring the significant reduced parameter space to determine an effective range of experimental parameters. We demonstrate in our model that the spectral cross talk between the ON and OFF transitions of the switch can be significantly reduced.

%Photoswitching from the ``ON state'' to the ``OFF state'' (the so-called ``forward reaction'') is often achieved by the same wavelengths of light as the ``reverse reaction" (``OFF'' to ``ON'') due to the overlapping absorption spectra of the states.

%\section{Background}\label{Sec:OG:BG}

   Cph8 is a recently developed phytochrome based optogenetic switch designed to control gene expression in E. Coli bacteria \cite{Levskaya2005}. Naturally occurring phytochromes are light sensing proteins in plants, algae, cyanobacteria, and other micro-organisms that harness light as an energy source \cite{Neff2000, Vierstra2000}. These organisms rely on phytochromes to respond to daily and seasonal changes to the light environment. In plants phytochromes control seed development and growth \cite{Franklin2010, Chen2011}, while in algae and cyanobacteria they act as primitive vision sensors to steer toward or away from light during phototaxis \cite{PHP:PHP1481,Garcia2000}. Phytochromes use the photo-isomerization of a covalently attached open-chain tetrapyrrole (bilin) chromophore to reversibly convert between two states in response to light. The chromophore cycles between a dark-stable, red-absorbing $P_R$ state and a meta-stable, far-red-absorbing $P_{FR}$ state, functioning as a molecular switch to regulate numerous responses to light intensity, color, duration, and direction \cite{Smith1995}.

\begin{figure}
  \centering
  \includegraphics{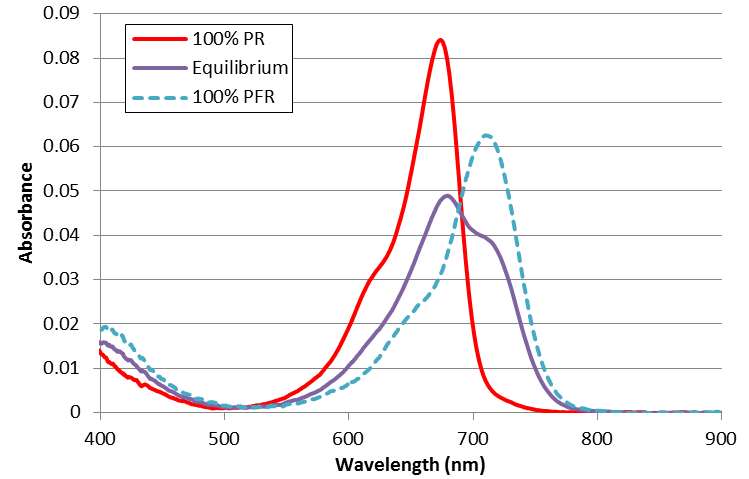}
  \caption[Cph8 Absorption]{Absorption spectra of the Cph8 sample.
   The red line shows the spectrum of the pure \pr state, obtained by maintaining the sample under saturated far-red illumination ($\lambda=725$ nm). The purple line is associated with a sample in the photoequilibrium mixed state of 65\% \pr and 35\% \pfr, obtained by maintaining the sample under saturated red illumination ($\lambda=650$ nm). From the difference of these spectra we can extract the pure \pfr spectrum, shown by the dashed blue line. The ultimate goal of this work is a full conversion (clean switching) to this projected pure $P_{FR}$ state.}
   \label{Fig:OG:Cph8Spec}
\end{figure}

Figure~\ref{Fig:OG:Cph8Spec} shows the absorption spectra of the Cph8 sample after exposure to red and far-red light until saturation of the photoswitching transition. After far-red exposure the switch has fully converted to the $P_R$ state, as this state does not appreciably absorb at wavelengths longer than 720 nm and there is no forward ($P_R\rightarrow P_{FR}$) photoswitching reaction.

The source of the spectral cross-talk is the overlap in the absorption spectra of the two chromophore states in the light-sensing module. To increase the dynamic range of the switch we must uncouple the optical interaction with the \pr state (which activates the forward photoswitching reaction) from the optical interaction with the \pfr state (which activates the reverse reaction). We have demonstrated theoretically \cite{ Li2002, Turinici2004, Li2005} and experimentally \cite{ Roth2009, Petersen2010, Rondi2011, Roslund2011, Rondi2012} that even molecules with spectra that are nearly functionally identical can be differentially excited by exploiting their unique, coherent excited state dynamics using ultrashort tailored pulsed light sources. To independently optically address the two states of the Cph8 protein complex we must understand the photo-isomerization dynamics of the PCB chromophore molecule.

As discussed above, the absorption of the $P_{FR}$ state extends to longer wavelengths than $P_R$, yielding near complete ($>$97\%) conversion from the $P_{FR}$ to the $P_R$ state (the reverse reaction) under constant far-red excitation. The opposite reaction is challenging due to spectral cross-talk: the red light excitation is absorbed by both the $P_R$ and $P_{FR}$ states.
Since we cannot solve the problem of spectral cross-talk by altering the absorption spectra, we must turn to an alternate set of optical interactions to find suitable controls to differentiate the states. While the ground state absorption of the $P_R$ and $P_{FR}$ states are similar and overlapping, the excited state dynamics are very different and open a new avenue for controlling the chromophores and differentiating the two states. Thus, we seek to exploit the distinct dynamics of the forward and reverse photo-isomerization reaction to halt the reverse reaction while minimizing the impact on the forward reaction.

The photo-isomerization of the PCB chromophore in both directions is initiated by photo-excitation of the molecule, immediately followed by very rapid coherent vibronic relaxation on the excited state surface \cite{Heyne2002}. This sequence is then followed by a portion of the excited molecules transitioning to a transient intermediate product state beginning the isomerization. The remaining excited state population returns to its respective initial ground state. In the forward isomerization reaction ($P_R\rightarrow P_{FR}$) this intermediate is LUMI-R; in the reverse ($P_{FR}\rightarrow P_R$) the intermediate is LUMI-F \cite{Bischoff2000,Bischoff2001}. These two transient intermediate species are unique to the forward and reverse reaction, respectively, and the subsequent relaxations progress along distinct, separate reaction coordinates.  The LUMI-R and LUMI-F transient intermediate products persist for several nanoseconds and trigger a series of large scale conformational reorganizations of the protein residues in the downstream domains to complete the photoswitching transformation between the $P_R$ and $P_{FR}$ states of the full Cph8 switch.

% \subsection{Description of the SDQ Mechanism}\label{Sec:OG:ExpDesc}

%\begin{figure}
%  \centering
%  \includegraphics[width=.65\linewidth]{ch-OG/Figs/ExpConcept_Schem}
%  \caption[Diagram of Experimental Concept]{Diagram of SDQ Experimental Concept: The Cph8 sample is exposed to the ultrashort Excitation and Depletion pulses, separated by $\uptau_{Delay}$. Both initial states of the switch are excited and the molecules undergo ultrafast relaxation toward the branching conical intersection to either the transient product states (LUMI-R and LUMI-F) or back to the initial ground states. There is good spectral overlap of the depletion pulse with the $P_{FR}$* excited state depletion spectrum (red arrow connects surfaces) and poor overlap with the $P_{R}$* depletion spectrum (red arrow too short to connect surfaces). A barrier on the $P_R$* excited surface slows the forward reaction. The combination of spectral overlap and timing permit the depletion pulse to coherently and selectively drive a larger portion of the $P_{FR}$* excited molecules back to $P_{FR}$ ground state, quenching the reverse reaction more than the forward. (the quenched molecule no longer follows the path of the grey dashed arrows)}\label{Fig:OG:ExpConcept}
%\end{figure}

In this theoretical simulation (SDQ Mechanism) selective stimulated emission is employed to coherently transfer PCB chromophores initially excited to the $P_{FR}$* state back to the \pfr ground state before isomerization can occur, while simultaneously allowing chromophores initially excited to the $P_{R}$* state to persist and undergo photoswitching to the $P_{FR}$ state. The sample is exposed to two ultrashort laser pulses: an excitation pulse and a depletion pulse.
The spectrum of the excitation pulse overlaps with the absorption spectra of both the $P_{R}$ and $P_{FR}$ states of the PCB chromophore. Exciting the molecule with the ultrashort pulse creates corresponding coherent wave packets from superpositions of vibrational levels on the electronic excited state surfaces of both forms of the chromophore.
These wave packets undergo dissimilar coherent dynamics as they relax from their initial Frank-Condon excitation states toward a coherent vibronic transition through a conical intersection either back to their initial ground state or to a transient intermediate photo-product state.
Before the molecules complete these dynamics, a second pulse arrives after a short, controlled delay ($\uptau_{Delay}$).
This depletion pulse has a central wavelength set to overlap more favorably with the stimulated emission spectrum of the $P_{FR}$* state than the $P_{R}$* state.
The combination of spectral overlap and timing permits the depletion pulse to selectively drive a larger portion of the $P_{FR}$* excited molecules back to the $P_{FR}$ ground state, preferentially slowing the reverse photo-isomerization reaction while allowing the forward reaction to continue transferring some portion of the \pr population to the \pfr state.
Those molecules remaining in either of the excited states after the depletion pulse continue their unperturbed dynamics, either isomerizing to the transient photo-product state or returning to their initial ground state.
%The relative final populations of the Cph8 switch in each state after being exposed to the pulsed excitation and depletion sequence are measured by absorption spectroscopy.

SDQ shares many similarities with Stimulated Emission Depletion (STED), a common super-resolution imaging technique used in microscopy. In STED microscopy the suppression of fluorescence from a ring-shaped outer spot provides contrast to an un-depleted inner spot, yielding an instrument response function with sub-diffraction limited spatial resolution.
%When using STED on specialized dyes, complete suppression of the fluorescence signal is possible.
In both STED and SDQ the goal is to fully de-populate an excited electronic state before the system can relax by an undesired pathway. STED microscopy provides improved spatial resolution, our goal is to provide improved spectral resolution. However, there is a critical difference between conventional STED of a fluorescent molecule and our proposed SDQ of the excited state of a photoswitch.
% that may prevent us from achieving the same degree of complete de-population of the excited state.The non-radiative photo-isomerization of the chromophore in the Cph8 sensor domain module is very different from emission of a dye or fluorescent protein. In fact, there is a class of Phytochrome-based fluorescent proteins designed specifically by disabling the photoswitching and blocking the isomerization of the chromophore in the sensor domain module, forcing the excited chromophore to decay back to the ground state by spontaneous emission \cite{Zhu2015, Shcherbakova2015, Shcherbakova2016 }.
%
In most fluorescent species the lifetime of the excited electronic state exceeds the lifetime of the coherent vibrational states by several orders of magnitude. In these species the molecule vibrationally relaxes to a stationary state on the electronic excited surface and a (relatively) long depletion pulse can interact with the population over an extended period, maximizing the population transferred to the ground state.
As discussed above,the vibrational and electronic excited state lifetimes are of similar duration for the chromophores,
and it is believed the molecule undergoes semi-coherent vibronic dynamics throughout the isomerization/relaxation (i.e., the initially coherent superposition persists while also interacting with the surrounding protein and solvent environments in ways which support the coherence or induce decoherence, the details of which are presently not well characterized \cite{Spillane2012,Mroginski2010}).
%from the instant the wave packet is formed on the excited state surface until it either returns to its initial ground state or transitions to the intermediate state that proceeds to isomerization and protein switching.
This impacts the potential effectiveness of the proposed SDQ mechanism in two ways.
First, because the excited state lifetime is so much shorter than a fluorescent species, the depletion pulse interacts with the molecule for less time, likely reducing the maximum achievable depletion.
Second, because the population is never in a stationary state on the excited surface, the depletion pulse must interact with a spectrally shifting wave packet, possibly negatively impacting the coherent transfer to the ground state.
Future technology developments should be able to overcome both of these obstacles by utilizing pulse shaping and feedback control to discover suitable pulse shapes for the excitation and depletion pulses that accommodate or exploit the excited state dynamics of the $P_{R}$* and $P_{FR}$* states in a maximally distinguishable fashion.
To demonstrate that even without optimally tailored pulse shaping the extremely rapid dynamics of the photo-isomerization would not prevent the SDQ mechanism from controlling the optogenetic switching of Cph8, we run simulations with a model system exhibiting comparable optical response and dynamics upon exposure to the dual-pulse excitation-depletion sequences.

In practical applications, optogenetic switches are expressed in large numbers in genetically targeted cells \cite{Mohanty2015}, not as isolated molecules. Optogenetic control of cellular processes involves the global photoswitching of a large population of protein switches from an arbitrary initial distribution of states to a maximally active or inactive population, restricted by the Maximal Dynamic Range (MDR) of the system and light exposure condition.  This will necessarily be a gradual transfer of population between states over many laser exposure iterations.
In particular, the low quantum efficiency of the photoswitching reaction (10-20\% in both reaction directions \cite{Yang2016}), as well as the fact that not all of the molecules will absorb photons in a single exposure, means that it would not be possible to transfer the entire population of a group of switches to a desired final state in one pulsed laser iteration.
The iterative transfer of population is made possible by the relative stability of both active and inactive states in phytochrome-based switches, allowing the state distribution to persist between laser exposure iterations. The \pfr state of Cph8 is meta-stable, reverting to the \pr state in the dark, but at such a slow rate (variable in different implementations and environments, from many seconds to many hours \cite{Olson2014}) that the state distribution can be said to be constant between laser pulses repeated at more than a few Hertz.
Each laser exposure iteration transfers a fraction of the proteins from one state to the other, with a per exposure fractional yield that is independent of the states' populations, defined exclusively by the characteristics of the light.
Because of the stability of the product state of the photoswitching reactions, we are able to transfer large populations of switches to the desired product state in many small steps.
This iterative transfer of population is important for a mechanism like SDQ, which acts by inhibiting an undesired reaction pathway rather than enhancing the rate of a desired pathway. The SDQ mechanism, while reducing the per-exposure fractional yield of the photoswitching reaction will none the less positively impact the final photoequilibrium after many exposures, which is the feature of interest for the practical implementation of controlled photoswitching.

In this work, the excitation and depletion pulse parameters will be varied to determine the wavelengths, powers, pulse durations, and delay timings which selectively deplete the $P_{FR}$* level most effectively, quenching the reverse reaction while minimally hindering the forward reaction.

%
%\singlespacing

\section*{\nohyphens{Rate Equation Model of Stimulated Emission Depletion Quenching}}\label{Sec:OG:REMSim}
%Control of photoswitching by Stimulated Depletion Quenching

%\doublespacing

We perform simulations using a Rate Equation Model (REM) to qualitatively assess how a system,(i.e., with optical characteristics and ultrafast dynamics corresponding to the Cph8 switch), responds to ultrashort-pulsed laser excitation and depletion.
 To this end, we first model how the system responds to a single excitation-depletion pulse sequence. These single exposure simulations generate a set of \emph{Characteristic Coefficients}, which describe the photoswitching reaction products for the entire set of possible initial conditions, defined exclusively by the laser exposure condition.
The products of the simulated single exposure photoswitching reaction are used to generate an Optical Transformation Matrix (OTM), which is a linear transformation from the initial to the final state population.
% defined solely by the characteristics of the laser pulse sequence.
The OTM can be used to simulate repeated switch exposures, thereby generating a mapping from the initial state to any final state of the system within the Maximal Dynamic Range (MDR) accessible by the associated laser exposure defined by the particular control parameter set.
We apply this iterative mapping to show how repeated exposures allow the $P_{FR}$ state to gradually reach a higher photoequilibrium enhanced by SDQ.
%\textcolor[rgb]{0.00,0.00,1.00}{The iterative transfer of population between the \pr and \pfr states over many exposures is an important and unique characteristic of the photoswitching reaction, as it defines an iterative mapping of the initial state to any final state within the dynamic range of the switch defined solely by the illumination of the sample. This iterative mapping is a function of the laser control parameters, and not the state of the system...}
Once the mechanics of the simulation are established we present a collection of results showing the functional dependence of the photo-reaction on the laser control parameters and locate the region of the parameter space that most effectively enhances the photoswitching dynamic range. These simulations are designed to provide a basis for subsequent experiments.

\subsection*{Simulation of Dual-Pulsed Excitation-Depletion Photoswitching}\label{Sec:OG:REMSingle}

\begin{figure}
  \centering
  \includegraphics[width=\linewidth]{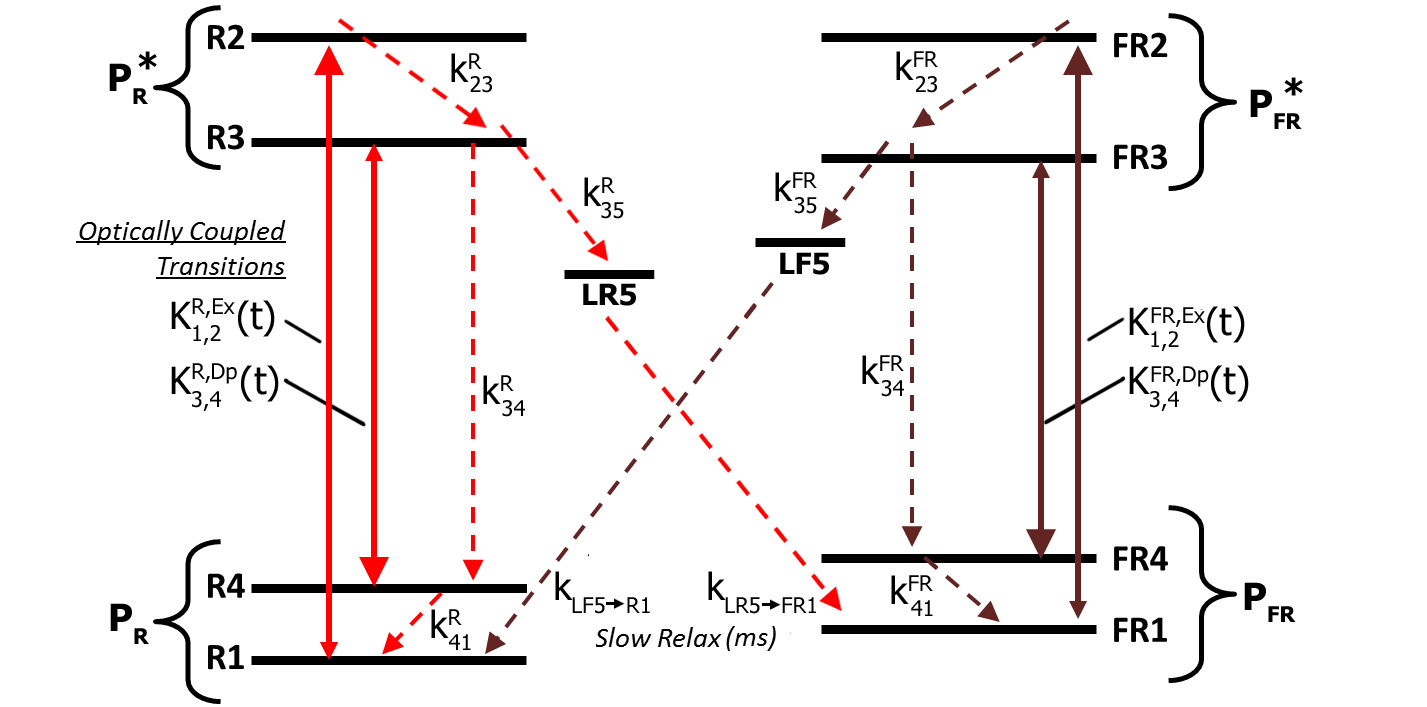}
  \caption[Rate Equation Model Schematic]{Schematic of relevant states and transitions for simulation of the photoswitching reaction by REM. Solid arrows represent optically coupled transitions between the electronic ground and excited states while dashed lines are spontaneous relaxations between states. The transitions are labeled by their rate coefficients: the constant values of the spontaneous transitions are collected in Table~\ref{Tab:OG:kijs} while the variable coefficients of the optically coupled transitions are functions of the time dependent laser controls as described by Eq~\ref{OG:RateMod:SpecRateConst} in the text. The intermediate states LUMI-R (LR5) and LUMI-F (LF5) do not interact, transferring their entire accumulated population to the ground level of the opposite chromophore state at the end of the photoswitching reaction.}
  \label{Fig:OG:UFModel}
\end{figure}

A schematic of the model system is shown in Figure~\ref{Fig:OG:UFModel}. The multi-dimensional potential energy surfaces of the $P_{R}$, $P_{R}$*,$P_{FR}$ and $P_{FR}$* states have each been collapsed to two discrete levels representing each surface, and a pair of levels represent the transient intermediate products LUMI-R=LR5 and LUMI-F=LF5, totaling ten levels.
Rather than describing discrete states separated by single-valued energy differences, the levels should be viewed as representing manifolds of molecular states. The optical responses and associated dynamics of the levels are defined by experimentally measured absorption and emission spectra and state lifetimes associated with time resolved measurements, as discussed further below.
The model supports excitation to a Frank-Condon point and subsequent relaxation in the electronic excited state followed by either isomerization to the intermediate product state or transition to a hot ground state by spontaneous relaxation or stimulated emission.
The populations of the intermediate states, LUMI-R and LUMI-F, do not cross or interact (i.e., the intermediate transition states during the forward isomerization from $P_{R}$ to $P_{FR}$ are not shared with the transient states of the reverse reaction); once isomerization begins it will continue unabated.
The branching ratio of the isomerization/spontaneous emission pathways is dictated by the relative transition rate constants, which are based on excited state level lifetimes from experimental measurements \cite{Kim20140417, Yang2016}.
%\footnote{Put in the following descriptions or just cite? ``... current literature: the $P_{R}$ rate constants were measured by Dual-Excitation pump-probe transient absorption spectroscopy \cite{ Kim20140417 } while the $P_{FR}$ rate constants were obtained by combined analysis of polarization resolved transient IR spectroscopy and femtosecond Stimulated Raman \cite{Yang2016 }''}
%\tcg{With further information about the excited state dynamics a more physically complete model incorporating a degree of coherence could be simulated. The results of the present rate equation model can be thought of as a conservative picture of what can be achieved by the SDQ mechanism, as it does not allow for pulse shaping of the excitation and depletion pulses to exploit coherence effects in the excited state dynamics. }
The rate constants associated with the spontaneous transitions in Figure~\ref{Fig:OG:UFModel} are collected in Table~\ref{Tab:OG:kijs}.
Using these parameters, the model reproduces the expected behavior under single-pulse excitation. In particular, when exposed to continuous red illumination ($\lambda=650$ nm) the switch reaches a mixed state equilibrium of $P_{FR}/P_{R}\simeq 65\%/35\%$ and fully resets to $>99\%P_{R}$ under continuous far-red illumination ($\lambda >725$ nm).

\begin{table}[b]
\centering
\small
  \begin{tabular}{ @{}m{1.2in}@{} @{}m{1.2in}@{} @{}m{1.15in}@{} @{}m{1.152in}@{} @{}m{1.3in}@{} }
	\toprule
	\thead{ Electronic \\ Relaxation } & \thead{ Fast \\ Isomerization } & \multicolumn{2}{c}{\thead{Vibrational Relaxation}} & \thead{Slow \\ Isomerization} \\
	\midrule
	$k_{3,4}^{R} = \nicefrac{1}{26 ps}$ & $k_{3,5}^{R} = \nicefrac{1}{61 ps}$ & $k_{2,3}^{R} = \nicefrac{1}{0.15 ps}$ & $k_{4,1}^{R} = \nicefrac{1}{0.15 ps}$ & $k_{LR5-FR1}=\nicefrac{1}{10^9 ps}$\\
    $k_{3,4}^{FR} = \nicefrac{1}{0.30 ps}$ & $k_{3,5}^{FR} = \nicefrac{1}{1.50 ps}$ & $k_{2,3}^{FR} = \nicefrac{1}{0.05 ps}$ & $k_{4,1}^{FR} = \nicefrac{1}{0.05 ps}$ & $k_{LF5\rightarrow R1} = 	\nicefrac{1}{10^9 ps}$\\
    \bottomrule
    \end{tabular}
  \caption{Static rate coefficients of REM. Coefficients are expressed as inverse lifetimes associated with values from the literature \cite{ Kim20140417, Yang2016}. }\label{Tab:OG:kijs}
\end{table}

Because details of the structure and coherent dynamics of the chromophore molecules in the phytochrome switching systems are still not fully understood, the coherent energy-time coupling properties of the molecule and the laser pulses are not contained in this kinetic model.
The high-dimensional, vibronically coupled potential energy surfaces are simplified to a reduced number of levels associated with the integrated spectral response of the molecule and linked by spontaneous incoherent transitions.
The pulsed lasers are characterized under the slowly varying envelope (SVE) approximation: the temporal dynamics of the pulse serves only to scale the fixed spectral intensity distribution, leaving the coherent spectral phase and fast oscillating component of the complex-valued electric field out of the calculation.
These simplifications allow us to simulate the system without making further assumptions about characteristics of the excited state dynamics that are not conclusively accepted.

\begin{figure}
  \centering
  \includegraphics[width=\linewidth]{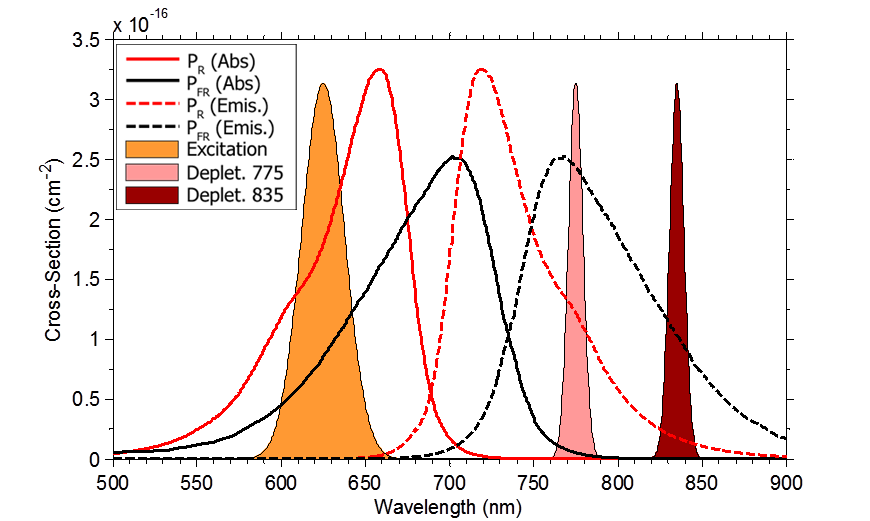}
  \caption[Spectra used in Rate Equation Model]{Absorption (solid lines) and stimulated Emission (dashed  lines) spectra of Cph8 in the $P_{R}$ state (red curves peaking at 665 nm and 725 nm, respectively) and the $P_{FR}$ state (black curves peaking at 710 nm and 770nm, respectively). Also plotted are spectra representing an excitation pulse centered at 625 nm and two different depletion pulses centered at 775 nm and 835 nm (orange, light red, and dark red shaded areas, respectively).} \label{Fig:OG:SimSpecs}
\end{figure}

Starting with the measured reference absorbance spectra of the Cph8 protein in the $P_{R}$ and $P_{FR}$ states, we can simulate or approximate the remaining features of the model.
%with a few reasonable assumptions and approximations.
The reference absorbance spectra are shown in Figure~\ref{Fig:OG:SimSpecs}, scaled to the peak molecular absorption cross-section. The absorption spectra in the literature~\cite{Lamparter2002} are reported as molar attenuation coefficients, but the calculations require the molecular cross-sections; the conversion between the two is simply: \(\sigma(\lambda)=\nicefrac{10^3ln(10)}{N_{Av}}\times\varepsilon(\lambda)\).
The peak molar attenuation coefficient of the \pr state is $\varepsilon_R (650\ nm)=85$ $(mM cm)\up{-1}$ and the corresponding value for the peak of the \pfr state is $\varepsilon_{FR} (705nm)=65.9$ $(mM cm)\up{-1}$, resulting in a peak absorption cross-section of $\sigma_R(650)=3.24\times10^{-16} cm^2$ and $\sigma_{FR}(705)=2.51\times10^{-16} cm^2$ for the \pr and \pfr states, respectively.
There is no measurement of the stimulated emission cross-section for the chromophore; however, we can represent the emission spectrum by reflecting the absorption spectrum across the zero phonon line \cite{MukamelNLO}. Comparing the absorption/emission spectra of similar species, notably the iRFP fluorescent protein species developed by mutation of the Cph1 photo-sensory module \cite{Zhu2015}, this symmetry is common and the assumption is justified. These simulated emission spectra are also plotted in Figure~\ref{Fig:OG:SimSpecs} using dashed lines.

We define the excitation and depletion pulses in terms of their real-valued spectral intensity, described here by a scaled Gaussian distribution. For the pulse \emph{p} = (\emph{Ex.} (excitation) or \emph{Dp.} (depletion)) centered at wavelength $\lambda_0^p$ and with full width at half max of $\Delta \lambda_p$, the real spectral intensity is:

\begin{equation}\label{Eq:OG:SpecIntensity}
  I_{p}(\lambda; \lambda_0^p, \Delta \lambda_p, \langle P_p \rangle,d_p)=\left(\dfrac{4\langle P_p \rangle}{\pi\cdot d_p^2}\right)\left(\dfrac{2\sqrt{ln2}}{\Delta \lambda_p \sqrt{\pi}}\right)e^{-({2\sqrt{ln2}\cdot (\lambda -\lambda_0^p)}/{\Delta \lambda_p})^2}
\end{equation}

\noindent The spectral distribution function is scaled by the average power $\langle P_{p} \rangle$ (in units of Watts) and beam waist diameter $d_{p}$ (in units of cm) to give the spectral intensity in (\nicefrac{W}{cm\up2}). The ``beam diameter'' is defined in these simulated pulses to describe the intensity in terms of experimentally measured and controlled parameters. The spectra of the excitation and depletion pulses are plotted with the molecular absorption and emission spectra in Figure~\ref{Fig:OG:SimSpecs} for an excitation pulse centered at 625 nm with width 30 nm and two separate depletion pulses with width 10 nm, one centered at 775 nm and the other at 835 nm.

The temporal intensity of the pulse is defined by a normalized unit area Gaussian function with variable delay $t_0^{p}$ and an adjustable pulse width $\Delta t_p$:

\begin{equation}\label{Eq:OG:TempIntensity}
  I_{pulse}(t; t_0^p, \Delta t_p,)=\left(\dfrac{2\sqrt{ln2}}{\Delta t_p \sqrt{\pi}}\right) e^{-\left({2\sqrt{ln2}\cdot (t-t_0^p)}/{\Delta t_p}\right)^2}
\end{equation}

\noindent 

With the reference absorption and emission spectra of the simulated Cph8 molecule and the modeled excitation and depletion pulses, we calculate the dynamic rate constants for the optical transitions in the molecule and simulate the time dependent populations of each level for both the $P_{R}$ and $P_{FR}$ states. The initial conditions are defined, with each state having a fraction of the total chromophore population in its lowest ground state:
\begin{align}\label{OG:RateMod:IC}
  [P_R]_{init} &= R1_0 \\
  [P_{FR}]_{init} &= FR1_0=(1-R1_0)
\end{align}
\noindent where $0 \leq R1_0 \leq 1$. The time dependent populations of the levels of the $P_{R}$ and $P_{FR}$ states are given by the solution to a set of coupled first order differential rate equations (see Figure~\ref{Fig:OG:UFModel} for reference to the notation).
The coupled differential equations describing the populations of the five levels associated with the \pr state (R1-LR5) are shown below in Equations~\ref{ER1}-\ref{ER5}. The five \pfr levels (FR1-LF5) are described by five more equations of the same form in Equations~\ref{EFR1}-\ref{EFR5}, exchanging appropriate rate constants and populations for the $P_{FR}$ levels.
The \pr and \pfr states are coupled by the transfer from level LR5 to FR1 (eqs. \ref{ER5},\ref{EFR1}) and LF5 to R1 (eqs. \ref{EFR5},\ref{ER1}).
The rate equations for the associated $P_R$ levels are:
\begin{subequations}
\label{EREM}
\begin{align}
\begin{split}
  \dfrac{d(R1(t))}{dt} ={}& [K_{1,2}^{R,Ex}(t)+K_{1,2}^{R,Dep}(t)]\cdot (R2(t)-R1(t))\\
  &\quad +(k_{4,1}^R\cdot R4(t))+(k_{LF5\rightarrow R1}\cdot LF5(t))
  \label{ER1}
  \end{split}\\
\begin{split}
  \dfrac{d(R2(t))}{dt} ={}& [K_{1,2}^{R,Ex}(t)+K_{1,2}^{R,Dep}(t)]\cdot (R1(t)-R2(t))\\
  &\quad -(k_{2,3}^R\cdot R2(t))
  \label{ER2}
  \end{split}\\
\begin{split}
  \dfrac{d(R3(t))}{dt} ={}& [K_{3,4}^{R,Ex}(t)+K_{3,4}^{R,Dep}(t)] (R4(t)-R3(t))\\
  &\quad +(k_{2,3}^R\cdot R2(t))-(k_{3,4}^R\cdot R3(t))-(k_{3,5}^R\cdot R3(t))
  \label{ER3}
  \end{split}\\
\begin{split}
  \dfrac{d(R4(t))}{dt} ={}& [K_{3,4}^{R,Ex}(t)+K_{3,4}^{R,Dep}(t)]\cdot (R3(t)-R4(t))\\
  &\quad +(k_{3,4}^R\cdot R3(t))-(k_{4,1}^R\cdot R4(t))
  \label{ER4}
  \end{split}\\
\dfrac{d(LR5(t))}{dt} ={}& (k_{3,5}^R\cdot R3(t))-(k_{LR5\rightarrow FR1} \cdot LR5(t)),
\label{ER5}
\end{align}
\end{subequations}

\noindent while the rate equations for the associated $P_{FR}$ levels are:
\begin{subequations}
\label{EREM_FR}
\begin{align}
\begin{split}
  \dfrac{d(FR1(t))}{dt} ={}& [K_{1,2}^{FR,Ex}(t)+K_{1,2}^{FR,Dep}(t)]\cdot (FR2(t)-FR1(t))\\
  &\quad +(k_{4,1}^FR\cdot FR4(t))+(k_{LR5\rightarrow FR1}\cdot LR5(t))
  \label{EFR1}
  \end{split}\\
\begin{split}
  \dfrac{d(FR2(t))}{dt} ={}& [K_{1,2}^{FR,Ex}(t)+K_{1,2}^{FR,Dep}(t)]\cdot (FR1(t)-FR2(t))\\
  &\quad -(k_{2,3}^{FR}\cdot FR2(t))
  \label{EFR2}
  \end{split}\\
\begin{split}
  \dfrac{d(FR3(t))}{dt} ={}& [K_{3,4}^{FR,Ex}(t)+K_{3,4}^{FR,Dep}(t)] (FR4(t)-FR3(t))\\
  &\quad +(k_{2,3}^{FR}\cdot FR2(t))-(k_{3,4}^{FR}\cdot FR3(t))-(k_{3,5}^{FR}\cdot FR3(t))
  \label{EFR3}
  \end{split}\\
\begin{split}
  \dfrac{d(FR4(t))}{dt} ={}& [K_{3,4}^{FR,Ex}(t)+K_{3,4}^{FR,Dep}(t)]\cdot (FR3(t)-FR4(t))\\
  &\quad +(k_{3,4}^{FR}\cdot FR3(t))-(k_{4,1}^{FR}\cdot FR4(t))
  \label{EFR4}
  \end{split}\\
\dfrac{d(LF5(t))}{dt} ={}& (k_{3,5}^{FR}\cdot FR3(t))-(k_{LF5\rightarrow R1} \cdot LF5(t)).
\label{EFR5}
\end{align}
\end{subequations}

In the rate equations above, the molecule moves from level \emph{i} to level \emph{j} via \emph{spontaneous transitions}, governed by rate constants $k_{i,j}^S$
or by \emph{optically coupled transitions} governed by coefficients $K_{i,j}^{S,pulse}(t)$, where the $S$=($R$ or $FR$) denotes the association with the $P_{R}$ or $P_{FR}$ initial state and pulse refers to either excitation (Ex) or depletion (Dep).
The transition coefficients have units $ps^{-1}$.
As stated earlier, the spontaneous transition coefficients are related to measured excited state dynamic lifetimes from the literature, and are collected in Table~\ref{Tab:OG:kijs}. The vibrational relaxation rate constants of both states ($k_{2,3}^S$ and $k_{4,1}^S$) are very high, associated with the very short vibrational relaxation lifetimes.
As such, levels 2 and 4 have fleeting populations serving primarily as intermediaries to levels 1 and 3.

The optically coupled transition rate coefficients ($K_{i,j}^{S,pulse}(t)$) are ``dynamic'', and vary with the intensity of the pulse temporal envelope. These coefficients are calculated individually for the excitation and depletion pulses. For the chromophore population in state \emph{S} = (\emph{R} or \emph{FR}), interacting with a \emph{pulse(p)}=(\emph{Ex}(Excitation) or \emph{Dep}(Depletion)) the time dependent rate of transition from level \emph{i} to \emph{j} is given by :
\begin{equation}\label{OG:RateMod:SpecRateConst}
  K_{i,j}^{S,pulse}(t)= I_{pulse}(t;t_0^p,\Delta t_p)\int{ \sigma_{i,j}^{S}(\lambda)\cdot I_{pulse}(\lambda)\cdot (\lambda/hc)d\lambda}
\end{equation}
Here, the rate coefficient has a temporally invariant spectral component given by the pulse photon flux (the pulse spectral intensity from equation~\ref{Eq:OG:SpecIntensity} divided by the energy per photon ($hc/\lambda$)) and the absorption/emission cross-section ($\sigma_{1,2}^{S}(\lambda)$ is the absorption spectrum of state $P_R$ or $P_{FR}$ in $cm^2$, while $\sigma_{3,4}^{S}(\lambda)$ is the stimulated emission spectrum).
The invariant spectral component is multiplied by the temporal intensity envelope of the pulse given in equation~\ref{Eq:OG:TempIntensity}. 

Combined, the product of the two functions in Eqation~\ref{OG:RateMod:SpecRateConst} gives a time dependent rate coefficient for the optically coupled transitions.

The final term in equation~\ref{ER1} ($k_{LF5\rightarrow R1}$) is the gain term from $LF5$ to $R1$ that mediates the {reverse} switching from $P_{FR}$ to $P_{R}$ by way of the transient intermediate LUMI-F.
The final term in equation~\ref{ER5} ($k_{LR5\rightarrow FR1}$) is the loss term from LR5 to $FR1$ that mediates the {forward} switching from $P_{R}$ to $P_{FR}$ by way of the transient intermediate LUMI-R.
These terms are present in the corresponding equations for the dynamics of the $P_{FR}$ levels LF5 (eq.~\ref{EFR5}) and FR1 (eq.~\ref{EFR1}), acting as loss and gain terms, respectively. The $k_{LR5\rightarrow FR1}$ and $k_{LF5\rightarrow R1}$ rates are much smaller than the other rate constants (more than \emph{eight} orders of magnitude). This distinction arises because the large scale conformation change of the protein domains associated with the transformation from the transient intermediate products to the final product states occurs on a completely different time scale than the initial photo-isomerization (i.e., respectively hundreds of microseconds compared to a few picoseconds).
When calculating the time dependent level populations, the $k_{LR5\rightarrow FR1}$ and $k_{LF5\rightarrow R1}$ rates are effectively zero, and the final transient product populations at the end of the simulated time must be handled by a separate calculation to give the final product state populations. %\mbox{([\pr]\sub{fin},[\pfr]\sub{fin})=(R1\sub{fin}}, FR1\sub{fin}).
The inclusion of the latter rates here serve to link the ODEs that describe the $P_R$ levels (equations~\ref{ER1}-\ref{ER5}) to the ODEs that describe the $P_{FR}$ levels (equations~\ref{EFR1}-\ref{EFR5}).
As stated earlier, the transfer from the intermediate states (LR5 and LF5) are 100\% efficient, with the entire accumulated population of LR5 and LF5 at the end of the simulated time being added to FR1 and R1, respectively, to complete the simulated photoswitching reaction with all switches in one of the two ground states.
%not shown, but equivalent in form to equations~\ref{ER1}-\ref{ER5}).

Solving the set of rate equations above gives the dynamic populations of the indicated levels. The two sets of five dynamically coupled ODEs were solved using Matlab's ODE45 solver, based on the explicit Runge-Kutta (4,5) integrator see~\cite{Dormand1980, Shampine1997}.
To avoid the differential equations becoming ``stiff'' the time domain of the calculation was broken into two intervals. The level populations are calculated over a sufficiently long time interval to allow the laser pulses to pass and the dynamic optically coupled rate coefficients to go to zero, taken as five pulse widths before the center of the first pulse and after the last pulse. After this time the equations are reduced to a simple exponential relaxation of the population excited during the optically driven dynamics back to the ground or transient levels. These relaxations can be analytically determined and the final product populations calculated directly once the results of the optically coupled transitions are computed.

With laser intensities of zero the dynamic rate equations for the \pr levels (equations~\ref{ER1}-\ref{ER5}) and the \pfr levels (equations~\ref{EFR1}-\ref{EFR5}) simplify to the following analytically solvable first order rate equations:
\begin{subequations}
\label{EqSteady}
\begin{align}
  \dfrac{d(R1(t))}{dt} ={}& k_{3,1}^R\cdot R3(t)
  \label{ESR1}\\
  \dfrac{d(R3(t))}{dt} ={}& -(k_{3,4}^R +k_{3,5}^R)\cdot R3(t)
  \label{ESR3}\\
  \dfrac{d(LR5(t))}{dt} ={}& k_{3,5}^R\cdot R3(t)
  \label{ESR5}\\
  \dfrac{d(FR1(t))}{dt} ={}& k_{3,1}^FR\cdot FR3(t)
  \label{ESFR1}\\
  \dfrac{d(FR3(t))}{dt} ={}& -(k_{3,4}^FR +k_{3,5}^FR)\cdot FR3(t)
  \label{ESFR3}\\
  \dfrac{d(LF5(t))}{dt} ={}& k_{3,5}^FR\cdot FR3(t)
  \label{ESFR5}
\end{align}
\end{subequations}

\noindent Here the transient levels 2 and 4 are removed, as any residual population of level 2 can be directly added to level 3, and level 4 can be bypassed without altering the final solution. The initial conditions of these reduced equations are set by the populations of the dynamic rate equation solutions at the end of the optically coupled time interval. For an intermediate level population at end of exposure time $T_1$ of [$R1_1, R3_1, LR5_1,F1_1, F3_1, LF5_1$]  the post-exposure exponential dynamics and steady-state populations at long time are calculated to be:

\begin{subequations}
\label{Eqsteadyt}
\footnotesize
\centering
\begin{align}
  R3(t)={}&R3_1\cdot e^{-(k_{3,1}^R +k_{3,5}^R)t}{} &\Rightarrow&R3_{S.S.}=0
  \label{ETR3}\\
  R1(t)={}&R1_1+\dfrac{k_{3,1}^R}{k_{3,1}^R+k_{3,5}^R}R3_1\left(1-e^{-(k_{3,1}^R+k_{3,5}^R)t}\right) & \Rightarrow&R1_{S.S.}=R1_1+\dfrac{k_{3,1}^R}{k_{3,1}^R+k_{3,5}^R}R3_1
  \label{ETR1}\\
  LR5(t) ={}& R5_1+\dfrac{k_{3,5}^R}{k_{3,1}^R+k_{3,5}^R}R3_1\left(1- e^{-(k_{3,1}^R +k_{3,5}^R)t}\right){} & \Rightarrow&LR5_{S.S.}= R5_1+\dfrac{k_{3,5}^R}{k_{3,1}^R+k_{3,5}^R}R3_1
  \label{ETR5}\\
  FR3(t) ={}& F3_1\cdot e^{-(k_{3,1}^{FR} +k_{3,5}^{FR})t}{} & \Rightarrow&F3_{S.S.}=0
  \label{ETFR3}\\
  FR1(t)={}&F1_1+\dfrac{k_{3,1}^{FR}}{k_{3,1}^{F}+k_{3,5}^{FR}}F3_1\left(1-e^{-(k_{3,1}^{FR}+k_{3,5}^{FR})t}\right) & \Rightarrow&F1_{S.S.}=F1_1+\dfrac{k_{3,1}^{FR}}{k_{3,1}^{FR}+k_{3,5}^{FR}}F3_1
  \label{ETFR1}\\
  LF5(t) ={}& F5_1+\dfrac{k_{3,5}^{FR}}{k_{3,1}^{FR}+k_{3,5}^{FR}}F3_1\left(1- e^{-(k_{3,1}^{FR} +k_{3,5}^{FR)}t}\right) & \Rightarrow&LF5_{S.S.}= F5_1+\dfrac{k_{3,5}^{FR}}{k_{3,1}^{FR}+k_{3,5}^{FR}}F3_1
  \label{ETFR5}
\end{align}
\end{subequations}

\noindent The ``long time'' at which the system reaches these steady state intermediate populations is $\sim$nanoseconds after excitation, which is orders of magnitude longer than the $\sim$ picosecond excited state relaxation/isomerization but still orders of magnitude shorter than the final LUMI-R/LUMI-F relaxation occuring over $\sim$ microseconds.

In the final step of the simulated photoswitching reaction, the accumulated transition state population is transferred to the switch's alternate ground state ($[P_{R}]_{fin}=R1_{S.S.}+LF5_{S.S.}$ and $[P_{FR}]_{fin}=FR_{S.S.}+LR5_{S.S.}$). This completes a single photoswitching reaction from a single dual-pulse excitation-depletion exposure iteration.
Next section describes how this final product state population distribution can subsequently be used as a new initial population condition, allowing the simulated photo-reaction to be repeated, replicating the gradual transfer of larger populations over multiple exposures.

% %% INTRODUCE THE CHARACTERISTIC COEFFICIENTS

For characterizing the effect of the depletion pulse in the selective quenching of the forward and reverse photoswitching reactions, it is useful to monitor not just the final product populations, but also the fraction of the population taking part in the forward and reverse reactions.
To this end, we define the transformation of the initial state population distribution to the final product state populations in terms of a set of ``\emph{characteristic coefficients}'': the forward and reverse Yield Coefficients ($Y_{R,FR}$ \& $Y_{FR,R}$) and the forward and reverse Quenching Coefficients ($Q_R$ \& $Q_{FR}$).
\begin{subequations}\label{eq:Sim:CharCoeffs}
  \begin{align}
    P_{R,f}={}&\left(1-(1-Q_R ) Y_{R,FR} \right)P_{R,f} + \left((1-Q_{FR} ) Y_{FR,R} \right)P_{FR,i}\\
    P_{FR,f}={}&\left((1-Q_R ) Y_{R,FR} \right)P_{R,i} + \left(1-(1-Q_{FR} ) Y_{FR,R} \right)P_{FR,i}
    \end{align}
\end{subequations}
\noindent Here, the per exposure yield in the forward direction, ${Y_{R,FR}}$, is defined as the fraction of molecules transferred from \pr to \pfr by the excitation pulse alone and the forward quenching coefficient, ${Q_R}$, is the relative change in forward yield brought on by the interaction with the depletion pulse; the corresponding reverse coefficients ${Y_{FR,R}}$ and ${Q_{FR}}$ are defined for the reverse reaction in the same way.
Importantly, these \emph{characteristic coefficients} are independent of the population distribution of the sample, being exclusively related to the variable experimental control parameters used to define and describe the excitation and depletion laser pulse pair.
In this way the \emph{characteristic coefficients} serve as a link between the control parameters and the optogenetic switch product output.
It is straightforward to see that the ratios of the steady-state isomerization product populations to the initial state populations are equal to the product of the yield and quench coefficients for the forward and reverse reactions, respecively:
\begin{subequations}\label{eq:Sim:LumiCCs}
  \begin{align}
    LR5_{S.S.}/R1_0={}&\left(1-(1-Q_R ) Y_{R,FR} \right)\\
    LF5_{S.S.}/FR1_0={}&\left(1-(1-Q_{FR} ) Y_{FR,R} \right)
    \end{align}
\end{subequations}
This parallel can serve as a useful means to extract information about the transient intermediate level populations without requiring ultrafast absorption measurements.
The Net Gain of \pfr for a single photoswitching event is the difference between the forward and reverse reaction products:
\begin{equation}\label{eq:OG:netgain}
  \Delta P_{FR} = (1-Q_{R})Y_{R,FR}\cdot P_{R}-(1-Q_{FR})Y_{FR,R}\cdot P_{FR}
\end{equation}

\begin{table}
  \centering
  \begin{tabular}{@{}l c c c c c}
    \toprule
    \thead{Laser Pulse}&\thead{$\mathbf{\langle P\rangle}$(mW)}&\thead{$\mathbf{\lambda_0}$ (nm)}&\thead{$\mathbf{\Delta \lambda}$ (nm)}&\thead{$\mathbf{\Delta t_{width}}$ (ps)}&\thead{$\mathbf{t_{Ex,Dep}}$ (ps)}\\
    \midrule
    Excitation&0.50&625&30&0.10&n/a \\
    Deplet. 775nm&10.0&775&10&0.140&0.080 \\
    Deplet. 835nm&10.0&835&10&0.140&0.080 \\
    \bottomrule
  \end{tabular}
  \caption[Model Laser Parameters]{Collective laser parameter settings for simulating the excitation and depletion pulses in Figures \ref{UFDeplete}, and \ref{Fig:MultiExp}.
  Average power ($\langle P\rangle$), central wavelength ($\lambda_0$) and spectral width ($\Delta \lambda$), temporal duration ($\Delta t_{width}$) and excitation-depletion delay ($t_{Ex,Dep}$) used in the calculation of pulse intensities with Eq.~\ref{Eq:OG:SpecIntensity} and Eq.~\ref{Eq:OG:TempIntensity}. These laser pulse intensities are exceptionally high, and used here only for illustration purposes. It is shown later that similar results can be achieved with lower pulse intensities. }
\label{Table:OG:LaserParam}
\end{table}

\begin{figure}
  \centering
  \includegraphics[width=\linewidth]{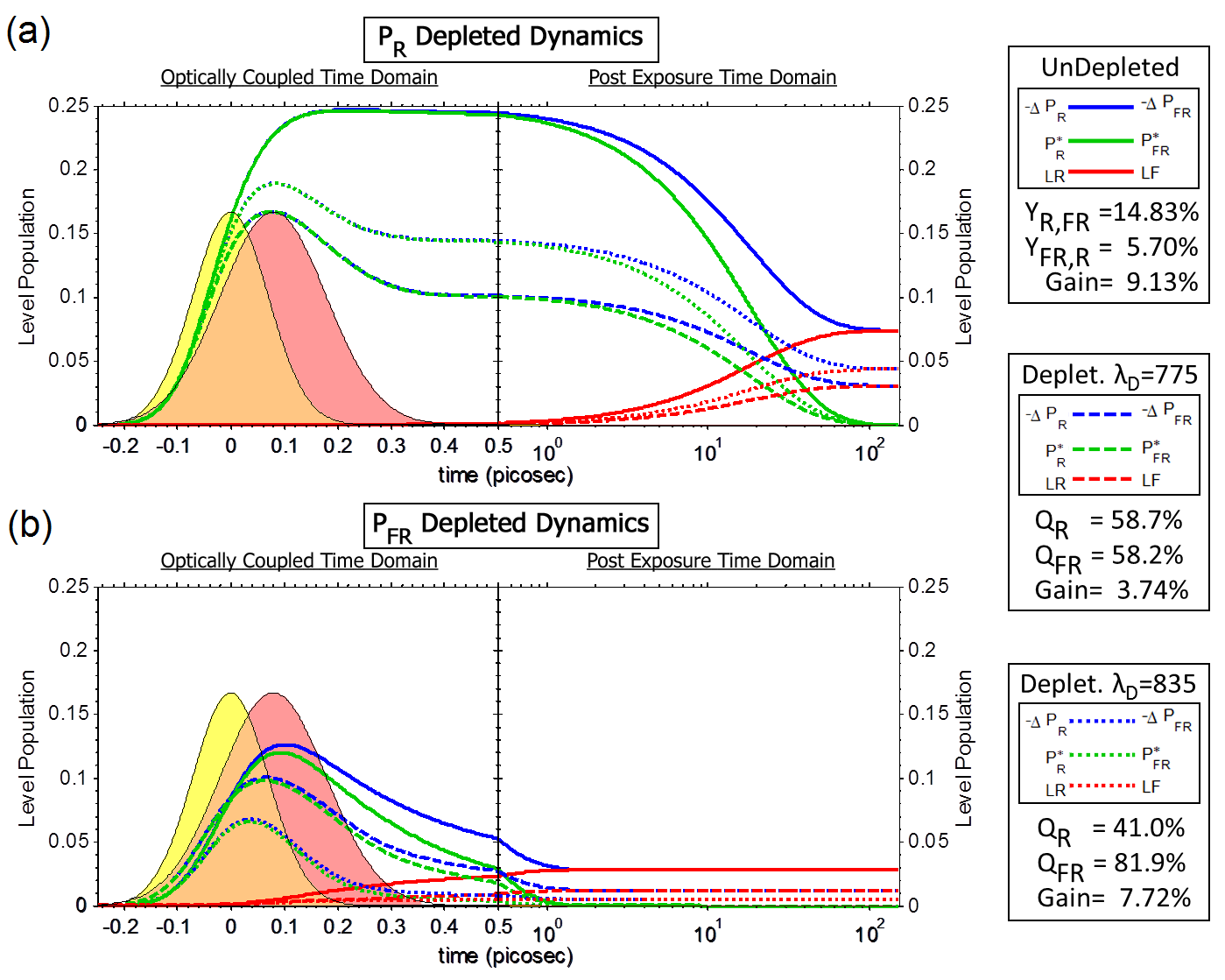}
  \caption[Modeled Ultrafast Depletion]{Comparing depleted dynamics of the $P_{R}$ (a) and $P_{FR}$ (b) states at two depletion wavelengths. The ground-state bleach ($-\Delta P_{R},-\Delta P_{FR}$), excited state ($P_{R}$*$,P_{FR}$*), and transient product (LR,LF) populations  of each state are plotted along with the excitation and depletion pulse envelopes (yellow and pink, respectively).
  The forward and reverse Yield and Quenching coefficients, and the final net Gain of \pfr are listed below the key for each exposure condition.
  The depletion at 835 nm with $Q_{FR}\sim(2\!\times\!Q_{R})$ effectively halts the reverse reaction while allowing the forward reaction to proceed. This produces a lower Gain per exposure, but will shift the photoequilibrium. The depletion at 775 nm quenches forward and reverse reactions equally.
  Laser parameter settings are listed in the text and Table~\ref{Table:OG:LaserParam}.} \label{UFDeplete}
\end{figure}

Let us illustrate the kinetics of the quenching process. Figure~\ref{UFDeplete} shows the altered dynamics after excited state quenching at two different depletion wavelengths.
To simplify the graphs, the figure shows the summed populations of the levels to represent the populations of the ground states ($R1+R4=P_{R}$, $FR1+FR4=P_{FR}$), excited states ($R2+R3=P_{R}$*, $FR2+FR3=P_{FR}$*), and intermediate product states (LR5=LR, LF5=LF).
%Stimulated depletion of the excited state reduces the transient products and recovers the ground state population.
The un-depleted level dynamics are plotted in the figure as solid lines. The dynamic level populations from two separate excitation-depletion exposures are shown, one with a central depletion wavelength of 775 nm (dashed traces) and the other centered at 835 nm (dotted traces), both with 10 nm bandwidth. The depletion pulses both have temporal full width at half max of 0.140 ps duration and arrive at a delay of 0.080 ps after the excitation pulse. The average power of both depletion beams is 10.0mW. This high depletion power is chosen to emphasize the effects of the stimulated depletion quenching in the figure.
It is shown later that similar final populations can be obtained with appropriate pulses of lower power.

The forward and reverse Yield and Quench coefficients and the \pfr Net Gain are shown on Figure~\ref{UFDeplete} by the key for each exposure condition. These values, as well as the final populations of the \pr and \pfr states, and the steady-state populations of the LUMI-R and LUMI-F transient states are collected in Table~\ref{Tab:ModResults}.
The final \pr and \pfr populations include the transfer of the accumulated population of the transient states to their final product ground states. The laser parameters associated with these results are collected in Table~\ref{Table:OG:LaserParam}.

\begin{table}
  \centering
  \begin{tabular}{@{}l c c c c c c c}
    \toprule
    \thead{Exposure\\ Condition\\{}}&\thead{Final \\ $\mathbf{P_{R}}$\\{}}&\thead{Final \\ $\mathbf{P_{FR}}$\\{}}&\thead{Net Gain\\ $\mathbf{\Delta P_{FR}}$\\{}} & \thead{LUMI-R \\ (${LR5_{S.S.}}$)\\{}} & \thead{LUMI-F \\ (${LF5_{S.S.}}$)\\{}} &\thead{Forward \\ Quench \\ $\mathbf{Q_R}$}&\thead{Reverse \\ Quench \\ $\mathbf{Q_{FR}}$}\\
    \midrule
    Un-Depleted&0.454&0.546&9.1\%&.0742&.0285& -- & -- \\
    Deplet.775nm&0.481&0.519&3.7\%&.0306&.019&58.7\%&58.2\%\\
    Deplet.835nm&0.461&0.539&7.7\%&.0438&.0052&41.0\%&81.9\%\\
    \bottomrule
  \end{tabular}
  \caption[Model Results: Yield and Depletion]{Collective results from exposure of an initial 50/50 mixed state to a single excitation-depletion pulse pair.
  The final populations are listed in the first two columns, followed by the Net Gain of the \pfr state. The intermediate steady-state populations of the transient isomerization products are next; followed by the Quenching coefficients, which give the percent change in the transient population with and without depletion.The forward/reverse Yield coefficients from the excitation pulse for all conditions are $Y_{R,FR}=14.83\%$ and $Y_{FR,R}=5.703\%$, respectively}
    \label{Tab:ModResults}
\end{table}

Quenching of the photo-isomerization is achieved at both depletion wavelengths for both states, as demonstrated by the drop in transient product states (levels LR5 and LF5, red in both graphs in Figure~\ref{UFDeplete}) and recovery of the initial ground states (levels R1 and FR1, blue in both graphs).
Looking at the values in Table~\ref{Tab:ModResults}, we see that SDQ \emph{lowers} the Net Gain ($\Delta P_{FR}$) per exposure for both depletion pulses, which would seem to run counter to our objective.
However, we will see that the $\Delta P_{FR}$ per exposure is less consequential than generating a significant difference between the Forward and Reverse Quenching Coefficients, $Q_{FR}\gg Q_{R}$.
The shorter wavelength depletion pulse ($\lambda_{0,D}= 780$ nm) quenches both $P_{R}$* and $P_{FR}$* excited states equally, $Q_{FR}\simeq Q_{R}$. This only slows the overall reaction, giving no selective control between the forward and reverse reaction directions.
%, giving a low \pfr Net Gain of 3.75\%,
The longer wavelength depletion pulse ($\lambda_{0,D}= 835$ nm) nearly completely depletes the $ P_{FR}$* excited state before much transient product can form, while the $ P_{R}$* excited state retains some population that continues the forward reaction.
This results in a lower Net Gain per exposure than the un-depleted excitation, $\Delta P_{FR}=7.72\%<9.13\%$, but nearly no loss of $P_{FR}$ to the reverse reaction.
It will be shown in the next section that this shift in the balance of the forward and reverse reactions enhances the maximum $P_{FR}$ population achieved at photoequilibrium over multiple exposures.

%$P_{FR}$*
These single exposure simulations show that with moderate laser parameters attainable experimentally it is reasonable to expect depletion of the excited state of the PCB chromophore in Cph8 before photo-isomerization occurs, thereby quenching the (\pfr$\rightarrow$\pr) reverse photoswitching reaction. The simulations further show that the extremely rapid dynamics of the $P_{FR}\rightarrow P_R$ photo-isomerization, rather than being a hindrance to control, allows selective depletion of the $P_{FR}$* excited state without complete depletion of the $P_{R}$* excited state, enabling favorable quenching of the reverse $P_{FR}\rightarrow P_R$ photoswitching and relative enhancement of the forward $P_{R}\rightarrow P_{FR}$ reaction.
Next, simulations of the system exposed to multiple laser exposures will show that the SDQ enhancement shifts the photoequilibrium, greatly increasing the dynamic range of the switch.

\subsection*{Multiple Exposures with an Optical Transformation Matrix}\label{Sec:OG:REMMulti}
%After simulating the dynamics of the system with a single excitation-depletion pulse sequence the final populations of the chromophore in states $[P_{R}]_{fin.}$ and $[P_{FR}]_{fin.}$ are then set as the new initial conditions of the rate \nohyphens{equations~\ref{ER1}-\ref{EFR5}} and the simulation could be repeated, incrementally transferring population between the $P_R$ and $P_{FR}$ states over many simulated exposures.
%This method of repeatedly solving the set of differential Rate Equations hundreds of times is not only slow, it is also prone to calculation errors due to round-off approximations and inaccuracies of the ODE solver. A better method is to recognize that this transfer of the initial state populations to the final state populations can be represented as a linear transformation of a vector of state populations by an \emph{Optical Transformation Matrix} (OTM) \textbf{T}:

The simulations of the last section determine the final population  ($P_{R,f}, P_{FR,f}$) after the initial population ($P_{R,i}, P_{FR,i}$) is exposed to a single Excitation-Depletion pulse sequence, defined by a set of laser control parameters, and by solving a set of differential rate \nohyphens{equations~\ref{ER1}-\ref{EFR5}}.
To determine the results of multiple exposures, rather than repeatedly solving the ODEs with new initial conditions at each iteration
it is more convenient to represent the transfer of the population of the system from an initial state to a final state as a linear transformation of a vector of state populations by an \emph{Optical Transformation Matrix} (OTM) \textbf{T}:

%\singlespacing
\begin{equation}\label{Eq_OTM}
 {\bf P_{f}}=\begin{bmatrix}
          P_{R,f} \\
          P_{FR,f} \\
        \end{bmatrix} =\begin{bmatrix}
                        T_{1,1} & T_{1,2} \\
                        T_{2,1} & T_{2,2}
                      \end{bmatrix} \cdot \begin{bmatrix}
          P_{R,i} \\
          P_{FR,i} \\
        \end{bmatrix}  = \textbf{T} \cdot {\bf P_{i}} \ .
\end{equation}

%\doublespacing

\noindent

The elements of the OTM are independent of the initial population distribution, as they are defined solely by the laser control parameters.
 To compute the values of the matrix elements we simulate a single iteration of the transformation using the rate equation ODEs, as described in the previous section, for two particular initial conditions and then solve for the matrix elements from the relationship between the initial and final populations. This reduces the number of times we must solve the rate equation system of ODEs from hundreds to just two.
In this way we can measure the SDQ enhancement, without having to wait for each permutation of the control parameters to reach photoequilibrium.
In the simulations, we solved the rate equations using the two initial conditions $\bigl[\begin{smallmatrix} 1 \\ 0 \\ \end{smallmatrix}\bigr]$ and $\bigl[ \begin{smallmatrix} 0 \\ 1 \\ \end{smallmatrix}\bigr]$, which provides the matrix elements directly:

%\singlespacing
\begin{equation}\label{Eq_OTMelements}
  \mathbf{T}\cdot \begin{bmatrix}
                             1 \\
                             0 \\
                           \end{bmatrix} = \begin{bmatrix}
                             T_{1,1} \\
                             T_{2,1} \\
                           \end{bmatrix} \textrm{, and }\mathbf{T}\cdot \begin{bmatrix}
                             0 \\
                             1 \\
                           \end{bmatrix} = \begin{bmatrix}
                             T_{1,2} \\
                             T_{2,2} \\
                           \end{bmatrix}
\end{equation}

%\doublespacing

\noindent highlighting the practical definition of the OTM elements: the off-diagonal elements $T_{1,2}$ and $T_{2,1}$ are the fractional gain per exposure between the two states.
%\pr to \pfr state (Forward Yield: $Y_{R,FR}$) and from the \pfr to the \pr state (Reverse Yield: $Y_{FR,R}$), respectively.
We could also refer to the definition of the \emph{characteristic coefficients} in Equation~\ref{eq:Sim:CharCoeffs} and write the OTM relation as:

%\singlespacing
\begin{equation}\label{eq:OG:OTMccs}
\small
 \begin{bmatrix}
          P_{R,f} \\
          P_{FR,f} \\
        \end{bmatrix} =\begin{bmatrix}
                        (1-(1-Q_{R})Y_{R,FR}) & (1-Q_{FR})Y_{FR,R} \\
                        (1-Q_{R})Y_{R,FR} & (1-(1-Q_{FR})Y_{FR,R})
                      \end{bmatrix} \begin{bmatrix}
          P_{R,i} \\
          P_{FR,i} \\
        \end{bmatrix}
\end{equation}

\begin{figure}
  \centering
  \includegraphics[width=.9\linewidth]{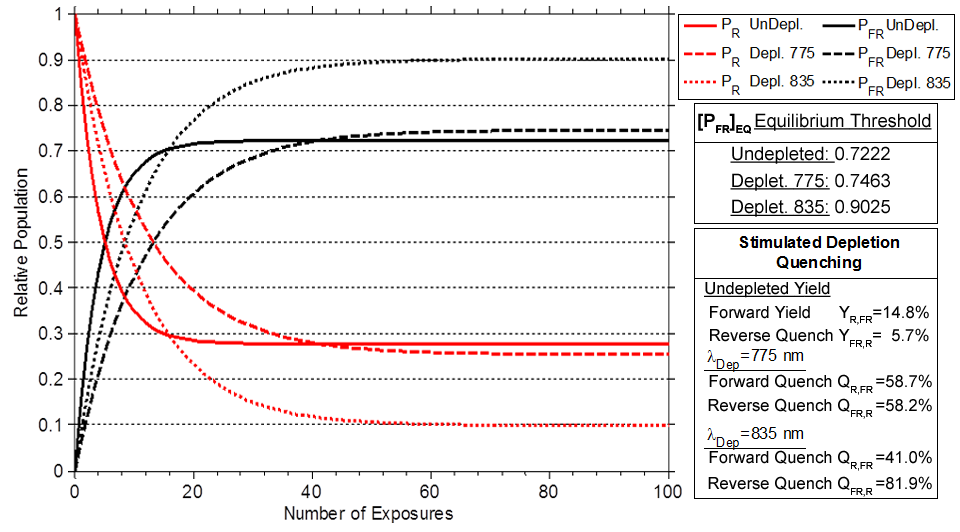}
  \caption[Modelled Multi-Exposure Accumulation of Photo-product]{Multi-exposure photo-isomerization yield with an un-depleted linear exposure, and an excitation-depletion pulse sequence exposure at two separate depletion wavelengths. In the first few exposures the faster reaction rate of the un-depleted exposure enables greater $P_{FR}$ generation, but after several exposures the quenching enhancement allows the sample exposed to the depletion at 830 nm to overcome the linear equilibrium limit and eventually reach a new equilibrium at $\sim$90.2\%. Laser parameters are listed in the text.}\label{Fig:MultiExp}
\end{figure}

%\doublespacing
\noindent To calculate the transfer of population from an arbitrary state distribution over a series $N$ of exposures the initial population vector is simply multiplied by, $\mathbf{T}^N$, the OTM raised to the number of exposures.
This iterative transfer of population over many laser exposures is possible because the population distribution of the system does not change between iterations when the period between laser pulses is less than the dark-relaxation time of the switch (minutes to hours for Cph8 \cite{Psakis2011}).
This gradual population transfer is plotted in Figure~\ref{Fig:MultiExp}. The system begins in the pure $P_{R}$ state $\bigl[\begin{smallmatrix} 1 \\ 0 \\ \end{smallmatrix}\bigr]$ and is examined under three different exposure conditions: un-depleted single-pulse excitation, and two sets of sequential excitation-depletion exposures with a depletion pulse centered at 775 nm and 830 nm.
The laser parameters for Figure~\ref{Fig:MultiExp}, collected in Table~\ref{Table:OG:LaserParam}, are the same as those used to plot Figure~\ref{UFDeplete}.
%The excitation pulse has average power 0.50 mW, spectral width of 30 nm centered at 625 nm, and temporal duration 100 fs. The depletion pulses both have average power 10 mW, spectral width of 10 nm, and temporal duration 140 fs. The excitation-depletion delay is 80 fs for both depletion pulses.
The photoequilibrium threshold (i.e., the maximum \pfr or minimum \pr population) under each exposure condition is annotated on the figure. Also noted are the forward and reverse Yield associated with the excitation pulse and the forward and reverse Quenching coefficients for each depletion pulse wavelength.
%the stimulated depletion quenching coefficient $Q_{i,f}$, the percent change in per pulse yield from state \emph{i} to state \emph{f} due to SDQ and the un-depleted yield: $Q_{i,f}= \left(\nicefrac{\left(Y_{i,f}^{(Ex)}-Y_{i,f}^{(ExDp)}\right)}{Y_{i,f}^{(Ex)}}\right)$.}Under un-Depleted red-light exposure the sample stabilizes at a photo-equilibrium threshold: 72.22\% $P_{FR}$ 27.78\% $P_{R}$.
When the sample is exposed to the excitation-depletion pulse pair, with the depletion pulse centered at 775 nm, the depletion cross-section is similar for both states of the chromophore, and the quenching parameters of the forward and reverse reaction are nearly identical ($Q_{R,FR} = 58.7\%, Q_{FR,R} = 58.2\%$). As a result, the depletion produces a small increase in the final equilibrium (74.63\% $P_{FR}$ 25.37\% $P_{R}$); thereby mainly serves to slow the reaction and increasing the number of exposures to reach 90\% of the equilibrium threshold ($N_{DEC}$) from 11 to 28 exposures.
When the central wavelength of the depletion pulse is 835 nm the photoswitching reaction continues rising past the linear threshold to a new enhanced equilibrium.
Initially, the slower reaction rate of the selectively quenched photoswitching produces less $P_{FR}$ than the standard linear excitation; however, after $\sim$20 exposures the quenched photoswitching reaction surpasses the cross-talk limited linear excitation threshold, achieving the final product population of 90.25\% $P_{FR}$ and 9.75\% $P_{R}$.
% with a set of depletion pulse parameters of 10 nm bandwidth at a central wavelength of 840 nm, a temporal pulse width of 0.14 ps, and an excitation-depletion delay of 0.06 ps.
This selective depletion significantly enhances the final yield of the desired $P_{FR}$ product and decreases the undesired $P_{R}$ product by a factor of 4 compared to the linear exposure limit.

Another feature of the iterative mapping formulation of the photoswitching process is that it is possible to solve for the associated eigenvectors and eigenvalues of the OTM:

%\singlespacing
\begin{equation}\label{Eq_Eigensols}
  \textbf{T}\cdot \textbf{v}_n = \lambda_n \textbf{v}_n
\end{equation}

%\doublespacing
\noindent which provide valuable information.
%For a simple 2x2 matrix we can directly calculate the eigensolutions. For a matrix
%
We know from the physics of the photoswitching reaction that one of the eigensolutions will have an eigenvalue of $1$ and be associated with the equilibrium state population. Once equilibrium is reached subsequent exposures no longer alter the population, so $\textbf{T}^N\cdot \textbf{v}_{eq}=1\cdot \textbf{v}_{eq}$. The set of two eigenvectors are linearly independent and span the associated space, any initial population vector can be expressed in this eigenbasis:

%%\singlespacing
%\begin{equation}\label{Eq_Eigenbasis}
%  \textbf{P}=\begin{bmatrix} P_{R}\\ P_{FR}\\ \end{bmatrix}=P_{R}\begin{bmatrix} 1\\ 0\\ \end{bmatrix}+P_{FR}\begin{bmatrix} 0\\ 1\\ \end{bmatrix}=c_{eq}\begin{bmatrix} R_{eq}\\ F_{eq}\\ \end{bmatrix}+c_{tr}\begin{bmatrix} R_{tr}\\ F_{tr}\\ \end{bmatrix}
%\end{equation}
%\singlespacing
\begin{equation}\label{Eq_Eigenbasis}
  \textbf{P}=\begin{bmatrix} P_{R}\\ P_{FR}\\ \end{bmatrix}=c_{eq}\begin{bmatrix} R_{eq}\\ F_{eq}\\ \end{bmatrix}+c_{tr}\begin{bmatrix} R_{tr}\\ F_{tr}\\ \end{bmatrix}
\end{equation}

%\doublespacing

\noindent and application of \textbf{T} to the population vector simply becomes:

%\singlespacing
\begin{equation}\label{Eq_Eigenbasis}
  \textbf{T}\cdot \textbf{P}=1\cdot c_{eq}\begin{bmatrix} R_{eq}\\ F_{eq}\\ \end{bmatrix}+ \lambda_{tr}\cdot c_{tr}\begin{bmatrix} R_{tr}\\ F_{tr}\\ \end{bmatrix}
\end{equation}
%\doublespacing

\noindent In this form it is clear that the second eigensolution is associated with the population that is transferred between states to take the initial population vector closer to the equilibrium eigenvector with each iteration of the transform.
The second eigenvalue $\lambda_{tr}$ is related to the number of exposures it takes to reach the equilibrium state.
From this eigenvalue we define the \emph{Decimation Count}, $N_{Dec}$, as  the number of exposures necessary to reduce the transitional population to less than 10\% of it’s initial value:

%\singlespacing
\begin{equation}\label{Eq_NDec}
  N_{Dec}=\dfrac{-1}{log_{10}(\lambda_{tr})}
\end{equation}
%\doublespacing

\noindent The decimation count is a useful metric of the effectiveness of a set of laser exposure parameters: a pulse pair that achieves a final equilibrium threshold of 100\% \pfr and 0 \% \pr but requires infinitely many pulses to reach this equilibrium is not practically useful.

Knowledge of the eigensolutions of the OTM provide a direct, simple means to calculate the photoequilibrium threshold and efficacy of a set of laser control parameters without having to \emph{calculate} or \emph{measure} the photoswitching product populations over many iterations until the system actually reaches the photoequilibrium.

\subsection*{Parameter Space Survey Results, Suppression of spectral cross talk}\label{Sec:OG:REMSurvey}

\begin{figure}
  \centering
  \includegraphics[width=.8\linewidth]{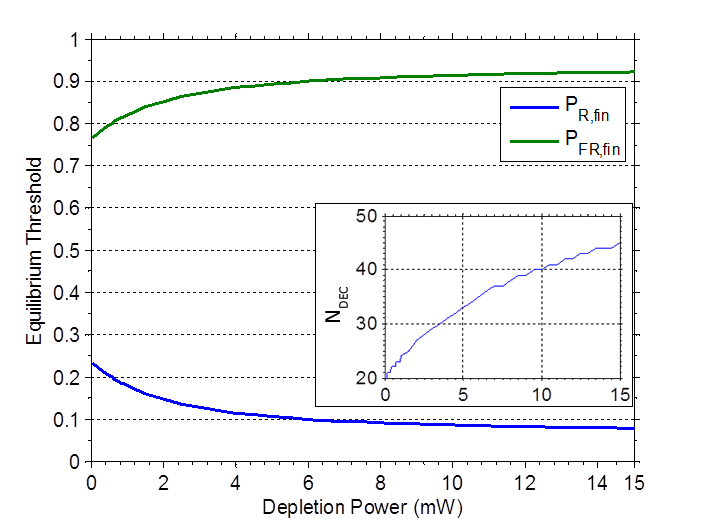}
  \caption[Power Dependence of the SDQ]{Dependence of the SDQ enhanced equilibrium population on the depletion pulse power; under fixed excitation power, spectrum, duration, and excitation-depletion delay. With these parameters the enhanced \pfr population surpasses 90\% with a depletion power of 6.5 mW. The inset shows that while increasing the power enhances the final equilibrium population, it also increases the decimation count $N_{DEC}$ by slowing the photoswitching reactions in both directions.}\label{DepPowScan}
\end{figure}

With the methods of calculating the photoswitching reaction product populations and final photoequilibrium threshold established in the previous sections, we survey the reduced space of laser control parameters that can be experimentally varied without complex pulse shaping to determine the dependence of the SDQ enhancement of the photoswitching reaction on these parameters independently and in concert.
Except where otherwise noted, the following surveys are conducted around a fixed starting point in the parameter space specified in Table~\ref{Tab:SurveyStart}.
The excitation and depletion pulse pair described in the table produces the highest SDQ enhancement accessible by varying the chosen control parameters, reaching an enhanced equilibrium population of $>90\%$ \pfr and $<10\%$ \pr with powers, spectra, and pulse durations accessible with reasonable laser sources.
Increasing the pulse peak intensities can enable higher equilibrium dynamic range, however this is achieved at the expense of slower reactions taking much longer to reach equilibrium, as shown later in Figure~\ref{DepPowScan}. Above an excitation power of 1 mW or a depletion power of 10 mW increasing the pulse power gives diminishing returns and enters an intensity range where the rate equation model used here will become increasingly inaccurate. While this location in the control space produces the best SDQ enhancement from a survey within this restricted parameter basis, it is emphasized that reaching full performance optimality will require more sophisticated quantum dynamics modeling and coherent pulse shaping (i.e., guided by optimal feedback control algorithms in the laboratory to address and exploit the complex coherent dynamics of the system).
The present results should be viewed as illustrative of the effectiveness of the SDQ mechanism for enhancing the dynamic range of the switch.

\begin{table}
  \centering
  \begin{tabular}{@{}l c c c c c}
    \toprule
    \thead{Laser Pulse}&\thead{$\mathbf{\langle P\rangle}$(mW)}&\thead{$\mathbf{\lambda_0}$ (nm)}&\thead{$\mathbf{\Delta \lambda}$ (nm)}&\thead{$\mathbf{\Delta t_{width}}$ (ps)}&\thead{$\mathbf{t_{Ex,Dep}}$ (ps)}\\
    \midrule
    Excitation&0.20&625&30&0.100& -- \\
    Depletion&6.5&833&10&0.140&0.080 \\
    \bottomrule
  \end{tabular}
  \caption[SDQ  Laser Parameters]{Best laser parameters for SDQ enhancement of Cph8 dynamic range based on a few-parameter survey of the control space.
  For this parameter set, SDQ enhanced the photoequilibrium to 90.25\% \pfr and 9.75\% \pr. Increasing the pulse powers can always increase the dynamic range, though at the expense of slower reactions and higher decimation count, $N_{DEC}$. Fully optimized control of the switches to their maximum dynamic range will require more elaborate quantum models and proper shaping of the laser pulse pairs in SDQ.}
\label{Tab:SurveyStart}
\end{table}

We first confirmed that the model behaves as expected in the single pulse excitation regime.

Because the absorption spectra of the chromophores is so much broader than the width of the excitation or depletion pulses, varying the spectral width of the excitation pulse has little effect on the photoswitching behavior. The molecule still fully resets to \pr under far-red illumination and reaches the same photoequilibrium under red illumination, increasing the spectral width only smears out the transition between these two regions of the molecular response.
Varying the pulse temporal duration similarly has essentially no effect on the single-pulse photoswitching reaction.
With the excitation pulse power fixed at 0.20 mW and the spectrum centered at 625 nm with a bandwidth of 30 nm, stretching the pulse from 20 fs to 2 ps changes the equilibrium population by $\sim2\%$.
Having confirmed that the model returns the expected results in the linear, single-pulse exposure conditions we can begin to test the dependence on the depletion pulse parameters.

%Next we scan over the excitation power to confirm that the system reaches the same equilibrium value regardless of excitation power (!!EXCEPT IT DOESN'T!!)
%\begin{figure}
%  \centering
%  \includegraphics[width=\linewidth]{ch-OG/Figs/REM/ExPscan}
%  \caption[Excitation power dependence of photoswitching]{Excitation power dependence of photoswitching}\label{ExWLScan}
%\end{figure}

\begin{figure}
  \centering
  \includegraphics[width=.8\linewidth]{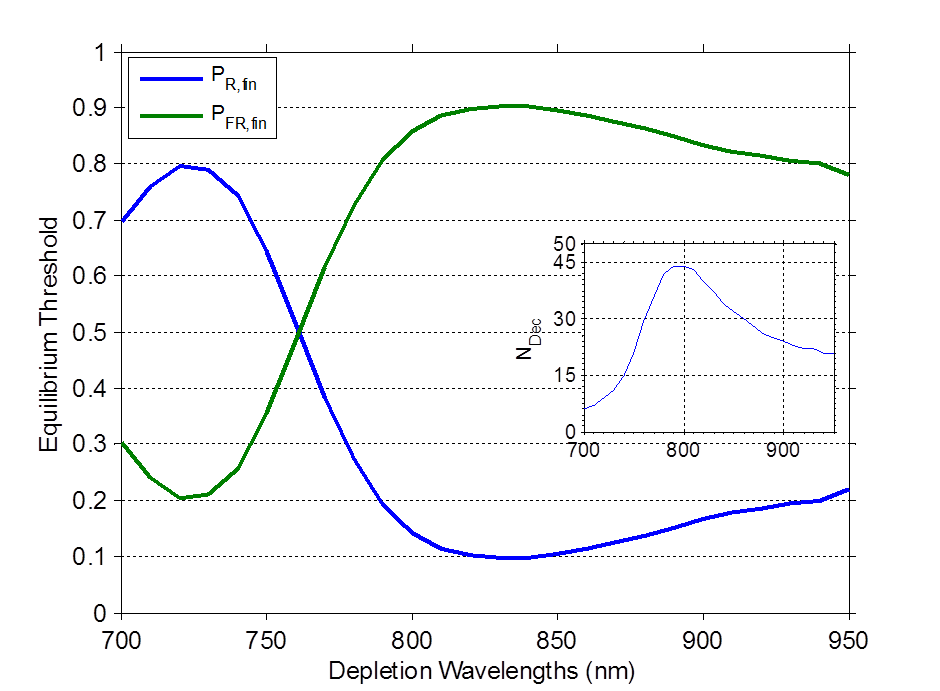}
  \caption[Depletion Wavelength Dependence the Stimulated Depletion Quenching]{Dependence of the SDQ enhanced equilibrium population of the \pr and \pfr states on the depletion wavelength, under fixed excitation power, depletion power, and pulse durations and delays. The optimum depletion wavelength is approximately 833 nm, yielding an enhanced final \pfr population of 90.25\%.}\label{DepWLScan}
\end{figure}

The depletion pulse central wavelength was scanned over a series of values to find the optimal depletion spectrum to maximize the final $P_{FR}$ product yield and minimize the unwanted \pr population.
Figure~\ref{DepWLScan} shows the dependence of the population as the depletion wavelength is scanned from 700-950 nm. The stimulated emission cross-section of both states is higher at shorter wavelengths, increasing the depletion quenching of the excited state, but reducing the selectivity. In addition, at $\lambda_{Dep}<$750 nm the \pfr ground state is also absorbing the depletion pulse at a non-vanishing rate, initiating the reverse reaction we are attempting to halt.
At longer wavelength there is an increased difference in the response of the states but lower overall depletion due to the falling cross-section. The maximum effective enhancement of the \pfr state is found between 830 - 840 nm, with the enhanced photoequilibrium of $[P_{FR}]_{EQ}=90.25\%$. Varying the spectral width of the depletion pulse has slightly more effect than varying the excitation spectral width, but again is the same as averaging over the features of the wavelength dependence. Without the use of complex pulse shaping, the depletion spectrum should be kept as narrow as possible while still being broad enough to support the desired temporal pulse characteristics.

Next we numerically investigate the power dependence of the SDQ. The depletion pulse power is varied while the spectrum is fixed at the optimal wavelength of 833 nm with a spectral width of 10 nm, the excitation pulse is 0.20 mW, 30 nm bandwidth centered at 625 nm, with a pulse duration of 100 fs. The depletion pulse duration is 140 fs and the excitation-depletion delay is 0.08 ps. Figure~\ref{DepPowScan} shows the power dependence of the SDQ enhanced populations. Even at low depletion power there is an increase in \pfr over the linear equilibrium of 70\%.
With these exposure parameters the SDQ enhanced \pfr population passes 90\% with a depletion pulse power of 6.5 mW.
Within this model, the enhancement of the \pfr state is likely not to reach 100\% \pfr, as there will always be a residual reverse reaction to produce a trace of \pr. As the depletion power increases the final photoequilibrium is shifted to higher \pfr population, however the higher depletion powers also slow the photoswitching reaction, thereby increasing the number of exposures necessary to reach this higher equilibrium state, as shown in the chart of the $N_{DEC}$ inset in Figure~\ref{DepPowScan}.
%Further, we saw in the single exposure simulations in Section~\ref{Sec:OG:REMSingle} that due to the ultrafast dynamics of the photo-isomerization it is not realistically possible to achieve 100\% depletion of the $P_{FR}$* excited state without also completely depleting the $P_{R}$* excited state, some small population is always able to begin isomerizing.
The maximum \pfr population achievable with a depletion power of 10 mW is 92.5\%; increasing the power to 50 mW this can be increased to 92.7\% Further increase in power gives only the marginal improvement while risking the dielectric breakdown. More sophisticated models or experimental operating conditions might bring us to the 100\% \pfr state.

\begin{figure}
  \centering
  \includegraphics[width=.95\linewidth]{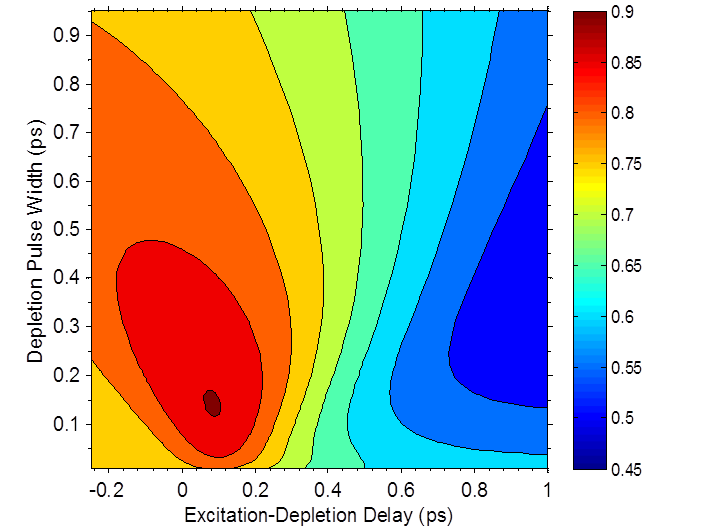}
  \caption[2-D scan of Depletion Temporal Parameters]{\uline{2-D scan of depletion pulse temporal parameters:} Dependence of the \pfr equilibrium population on the excitation-depletion pulse delay and depletion pulse duration. Depletion spectrum is set at the previously determined optimal value of 10 nm at 830 nm central frequency. The peak value of the enhanced \pfr photoequilibrium is 90.35\%  at a delay of 0.08 ps and pulse duration of 0.14 ps.}\label{Fig:OG:Temp2D}
\end{figure}

Finally we vary the temporal parameters of the depletion pulse. In general, increasing the depletion pulse duration (while maintaining the same instantaneous intensity) should increase the depletion of an excited state; the longer the depletion pulse interacts with the system the higher the probability it will trigger the transition to the ground state by stimulated emission. However, maximizing the depletion of the $P_{FR}^\ast$ state by lengthening the pulse will also lead to increased depletion of the slower to develop $P_{R}^\ast$ state, eliminating the selectivity of the quenching enhancement. The optimal pulse duration and delay will balance these opposing demands, interacting with the system as long as possible while still maintaining a selective enhancement. Beyond that, the gains of quenching the reverse reaction are outweighed by the losses caused by preventing the forward reaction.
To find the optimal depletion delay and pulse duration, these two parameters are scanned in concert at the optimal central wavelength of 833 nm with a depletion power of 6.5 mW. This 2-dimensional scan is shown in Figure~\ref{Fig:OG:Temp2D}. The peak of the surface occurs at a delay of 0.08 ps and pulse duration of 0.14 ps with an equilibrium \pfr population of 90.25\%, and the population of \pr of 9.75\%. In the linear, weak field excitation regime in is impossible to reduce the concentration of \pr state below 23\%. Thus, SDQ control allows us to significantly suppress the spectral cross talk. 

It is interesting to note that significant quenching enhancement can be achieved with the depletion pulse arriving before the excitation pulse if the width of the depletion pulse is greater than the negative delay. This allows the depletion pulse to interact at high intensity with the excited \pfr state as early as possible, without waiting for the leading edge of the depletion pulse to ramp up. This also helps to have the depletion pulse drop away before the excited \pr state fully develops. At higher power, the pulse can be moved to shorter or more negative delays, as the power on the trailing edge is high enough to overcome wasting the higher peak intensity before the population arrives at the lower level of the electronic excited state. Raising the power to 10 mW increases the equilibrium \pfr population to 92.5\%, but also shifts the optimal depletion pulse to 10 femtoseconds shorter delay and 25 fs longer duration to compensate. At a power of 10 mW it is possible to achieve higher than 90\% \pfr with zero delay pulses.

\section*{Conclusions}\label{Sec:OG:Conclusions}

%%\doublespacing

The results presented above demonstrate the capability to exploit stimulated depletion of the chromophore excited state to quench the photo-isomerization reaction before it takes place, diminishing the unwanted reverse photoswitching of the Cph8 optogenetic switch.
Further, by choosing optimal spectral and temporal characteristics of the excitation and depletion pulses, it is possible to selectively deplete the different chromophore states by exploiting their unique excited state dynamics and spectral response.

We developed a simple, flexible model of the Cph8 photoswitching system that can be adapted to the simulation of a number of alternate optogenetic switching systems or simultaneous control of multiple optogenetic switches.
The model is based on incoherent rate equation kinetics of the two switch states, and incorporates experimentally measured optical and dynamical characteristics of the molecular switch to simulate it's optically driven dynamics.
The model reproduces the expected steady-state photoequilibrium population distribution under saturated linear illumination with red or far-red light.
The predictions of this model are to be tested in upcoming publications.

%It is possible the model is incorrect at shorter delays where the excitation and depletion pulses overlap with each other and the earliest coherent dynamics of the switch.

%\singlespacing

Unconstrained shaping of both the excitation and depletion pulses in unison would enable the excitation of highly structured and unique excited state wave-packets that would undergo tailored unique dynamics, which the matching structure of the depletion pulse could follow in real time. This would allow for the most elaborate control over the molecules and yield true Optimal Dynamic Discrimination (ODD) of the two states of the chromophore in the optogenetic switches and full independent control of the optogenetic switching system.
Once a single switch can be reliably and independently controlled over its maximum dynamic range by non-linear photoswitching the next step would be simultaneous (multiplexed) optimal selective non-linear photoswitching of multiple optogenetic switches. The same unique excited state dynamics that enable ODD to differentially excite and deplete the two states of a single chromophore will be exploited to differentiate two similar optogenetic switches.

%% ==========================================================================
\singlespacing

\bibliographystyle{unsrt}
\bibliography{bib_thesis}

\begin{thebibliography}{10}

\bibitem{Deisseroth2011}
Karl Deisseroth.
\newblock Optogenetics.
\newblock {\em Nat Meth}, 8(1):26--29, Jan 2011.
\newblock Commentary.

\bibitem{Pastrana2011}
Erika Pastrana.
\newblock Optogenetics: controlling cell function with light.
\newblock {\em Nat Meth}, 8(1):24--25, Jan 2011.
\newblock Primer.

\bibitem{Fenno2011}
Lief Fenno, Ofer Yizhar, and Karl Deisseroth.
\newblock The development and application of optogenetics.
\newblock {\em Annual Review of Neuroscience}, 34(1):389--412, 2011.
\newblock PMID: 21692661.

\bibitem{zhang2011microbial}
Feng Zhang, Johannes Vierock, Ofer Yizhar, Lief~E Fenno, Satoshi Tsunoda, Arash
  Kianianmomeni, Matthias Prigge, Andre Berndt, John Cushman, J{\"u}rgen Polle,
  et~al.
\newblock The microbial opsin family of optogenetic tools.
\newblock {\em Cell}, 147(7):1446--1457, 2011.

\bibitem{hegemann2013channelrhodopsins}
Peter Hegemann and Georg Nagel.
\newblock From channelrhodopsins to optogenetics.
\newblock {\em EMBO molecular medicine}, 5(2):173--176, 2013.

\bibitem{Pudasaini2015}
Ashutosh Pudasaini, Kaley~K. El-Arab, and Brian~D. Zoltowski.
\newblock Lov-based optogenetic devices: light-driven modules to impart
  photoregulated control of cellular signaling.
\newblock {\em Front Mol Biosci}, 2:18, May 2015.
\newblock 25988185[pmid].

\bibitem{Levskaya2005}
Anselm Levskaya, Aaron~A. Chevalier, Jeffrey~J. Tabor, Zachary~Booth Simpson,
  Laura~A. Lavery, Matthew Levy, Eric~A. Davidson, Alexander Scouras, Andrew~D.
  Ellington, Edward~M. Marcotte, and Christopher~A. Voigt.
\newblock Synthetic biology: Engineering escherichia coli to see light.
\newblock {\em Nature}, 438(7067):441--442, Nov 2005.

\bibitem{Burgie2014Rev}
E.~Sethe Burgie and Richard~D. Vierstra.
\newblock Phytochromes: An atomic perspective on photoactivation and signaling.
\newblock {\em The Plant Cell}, 26(12):4568--4583, 2014.

\bibitem{2018arXiv181011432Q}
Z.~{Quine}, A.~{Goun}, K.~{Gerhardt}, J.~{Tabor}, and H.~{Rabitz}.
\newblock {Coherent Control of Optogenetic Switching by Stimulated Depletion
  Quenching}.
\newblock {\em ArXiv e-prints}, October 2018.

\bibitem{Neff2000}
Michael~M. Neff, Christian Fankhauser, and Joanne Chory.
\newblock Light: an indicator of time and place.
\newblock {\em Genes and Development}, 14(3):257--271, 2000.

\bibitem{Vierstra2000}
Richard~D. Vierstra and Seth~Jon Davis.
\newblock Bacteriophytochromes: new tools for understanding phytochrome signal
  transduction.
\newblock {\em Seminars in Cell and Developmental Biology}, 11(6):511--521,
  2000.

\bibitem{Franklin2010}
Keara~A. Franklin and Peter~H. Quail.
\newblock Phytochrome functions in arabidopsis development.
\newblock {\em Journal of Experimental Botany}, 61(1):11--24, 2010.

\bibitem{Chen2011}
Meng Chen and Joanne Chory.
\newblock Phytochrome signaling mechanisms and the control of plant
  development.
\newblock {\em Trends in Cell Biology}, 21(11):664--671, 2011.

\bibitem{PHP:PHP1481}
Brita Fiedler, Thomas Börner, and Annegret Wilde.
\newblock Phototaxis in the cyanobacterium synechocystis sp. pcc 6803: Role of
  different photoreceptors.
\newblock {\em Photochemistry and Photobiology}, 81(6):1481--1488, 2005.

\bibitem{Garcia2000}
M.~Garc{\'i}a-Dom{\'i}nguez, M.~I. Muro-Pastor, J.~C. Reyes, and F.~J.
  Florencio.
\newblock Light-dependent regulation of cyanobacterial phytochrome expression.
\newblock {\em J Bacteriol}, 182(1):38--44, Jan 2000.
\newblock 0791[PII].

\bibitem{Smith1995}
Smith Harry.
\newblock Physiological and ecological function within the phytochrome.
\newblock {\em Annual Reviews}, 46:289--315, 1995.

\bibitem{Li2002}
Baiqing Li, Gabriel Turinici, Viswanath Ramakrishna, and Herschel Rabitz.
\newblock Optimal dynamic discrimination of similar molecules through quantum
  learning control.
\newblock {\em J. Phys. Chem. B}, 106(33):8125--8131, 2002.

\bibitem{Turinici2004}
Gabriel Turinici, Viswanath Ramakhrishna, Baiqing Li, and Herschel Rabitz.
\newblock Optimal discrimination of multiple quantum systems: controllability
  analysis.
\newblock {\em Journal of Physics A: Mathematical and General}, 37(1):273,
  2004.

\bibitem{Li2005}
Baiqing Li, Herschel Rabitz, and Jean-Pierre Wolf.
\newblock Optimal dynamic discrimination of similar quantum systems with time
  series data.
\newblock {\em J. Chem. Phys.}, 122:154103, 2005.

\bibitem{Roth2009}
Matthias Roth, Laurent Guyon, Jonathan Roslund, V\'eronique Boutou, Francois
  Courvoisier, Jean-Pierre Wolf, and Herschel Rabitz.
\newblock Quantum control of tightly competitive product channels.
\newblock {\em Phys. Rev. Lett.}, 102:253001, 2009.

\bibitem{Petersen2010}
Jens Petersen, Roland Mitri\ifmmode~\acute{c}\else \'{c}\fi{}, Vlasta
  Bona\ifmmode \check{c}\else \v{c}\fi{}i\ifmmode \acute{c}\else~\'{c}\fi{}
  Kouteck\'y, Jean-Pierre Wolf, Jonathan Roslund, and Herschel Rabitz.
\newblock How shaped light discriminates nearly identical biochromophores.
\newblock {\em Phys. Rev. Lett.}, 105:073003, 2010.

\bibitem{Rondi2011}
A.~Rondi, D.~Kiselev, S.~Machado, J.~Extermann, S.~Weber, L.~Bonacia, J.-P.
  Wolf, J.~Roslund, M.~Roth, and H.~Rabitz.
\newblock Discriminating biomolecules with coherent control strategies.
\newblock {\em Chimia}, 65:346, 2011.

\bibitem{Roslund2011}
Jonathan Roslund, Matthias Roth, Laurent Guyon, Véronique Boutou, Francois
  Courvoisier, Jean-Pierre Wolf, and Herschel Rabitz.
\newblock Resolution of strongly competitive product channels with optimal
  dynamic discrimination: Application to flavins.
\newblock {\em J. Chem. Phys.}, 134(3):034511, 2011.

\bibitem{Rondi2012}
A~Rondi, L~Bonacina, a~Trisorio, C~Hauri, and J-P Wolf.
\newblock Coherent manipulation of free amino acids fluorescence.
\newblock {\em PCCP}, 14(26):9317--22, July 2012.

\bibitem{Heyne2002}
Karsten Heyne, Johannes Herbst, Dietmar Stehlik, Berta Esteban, Tilman
  Lamparter, Jon Hughes, and Rolf Diller.
\newblock Ultrafast dynamics of phytochrome from the cyanobacterium
  synechocystis, reconstituted with phycocyanobilin and phycoerythrobilin.
\newblock {\em Biophysical Journal}, 82:1004--1016, 2002.

\bibitem{Bischoff2000}
Mark Bischoff, Gudrun Hermann, Sabine Rentsch, Dietmar Strehlow, Stefan Winter,
  , and Haik Chosrowjan.
\newblock Excited-state processes in phycocyanobilin studied by femtosecond
  spectroscopy.
\newblock {\em The Journal of Physical Chemistry B}, 104(8):1810--1816, 2000.

\bibitem{Bischoff2001}
Mark Bischoff, Gudrun Hermann, Sabine Rentsch, and Dietmar Strehlow.
\newblock First steps in the phytochrome phototransformation: A comparative
  femtosecond study on the forward (pr to pfr) and back reaction (pfr to pr).
\newblock {\em Biochemistry}, 40(1):181--186, 2001.
\newblock PMID: 11141069.

\bibitem{Spillane2012}
Katelyn~M. Spillane, Jyotishman Dasgupta, and Richard~A. Mathies.
\newblock Conformational homogeneity and excited-state isomerization dynamics
  of the bilin chromophore in phytochrome cph1 from resonance raman
  intensities.
\newblock {\em Biophysical Journal}, 102(3):709 -- 717, 2012.

\bibitem{Mroginski2010}
Steve Kaminski and Maria~Andrea Mroginski.
\newblock Molecular dynamics of phycocyanobilin binding bacteriophytochromes: A
  detailed study of structural and dynamic properties.
\newblock {\em The Journal of Physical Chemistry B}, 114(50):16677--16686,
  2010.
\newblock PMID: 21126042.

\bibitem{Mohanty2015}
Samarendra~K. Mohanty and Vasudevan Lakshminarayananan.
\newblock Optical techniques in optogenetics.
\newblock {\em Journal of Modern Optics}, 62(12):949--970, 2015.

\bibitem{Yang2016}
Yang Yang, Karsten Heyne, Richard~A. Mathies, and Jyotishman Dasgupta.
\newblock Non-bonded interactions drive the sub-picosecond bilin
  photo-isomerization in the pfr state of phytochrome cph1.
\newblock {\em ChemPhysChem}, 17(3):369--374, 2016.

\bibitem{Olson2014}
Evan~J. Olson, Lucas~A. Hartsough, Brian~P. Landry, Raghav Shroff, and
  Jeffrey~J. Tabor.
\newblock Characterizing bacterial gene circuit dynamics with optically
  programmed gene expression signals.
\newblock {\em Nat Meth}, 11(4):449--455, Apr 2014.
\newblock Article.

\bibitem{Kim20140417}
Peter~W. Kim, Nathan~C. Rockwell, Shelley~S. Martin, J.~Clark Lagarias, and
  Delmar~S. Larsen.
\newblock Dynamic inhomogeneity in the photodynamics of cyanobacterial
  phytochrome cph1.
\newblock {\em Biochemistry}, 53(17):2818--2826, 2014.
\newblock PMID: 24742290.

\bibitem{Lamparter2002}
Tilman Lamparter, Norbert Michael, Franz Mittmann, and Berta Esteban.
\newblock Phytochrome from agrobacterium tumefaciens has unusual spectral
  properties and reveals an n-terminal chromophore attachment site.
\newblock {\em Proceedings of the National Academy of Sciences},
  99(18):11628--11633, 2002.

\bibitem{MukamelNLO}
Shaul Mukamel.
\newblock {\em Principles of Nonlinear Optical Spectroscopy}.
\newblock Oxford University Press, New York, 1995.

\bibitem{Zhu2015}
J.~Zhu, DM~Shcherbgakova, Y~Hontani, VV~Verkhush, and JTM Kennis.
\newblock Ultrafast excited-state dynamics and fluorescence deactivation of
  near-infrared fluorescent proteins engineered from bacteriophytochromes.
\newblock {\em Scientific Reports}, 5, 2015.

\bibitem{Dormand1980}
J.~R. Dormand and P.~J. Prince.
\newblock A family of embedded runge-kutta formulae.
\newblock {\em Journal of Computational Applied Mathematics}, 6:19--26, 1980.

\bibitem{Shampine1997}
L.~F. Shampine and M.~W. Reichelt.
\newblock The matlab ode suite.
\newblock {\em SIAM Journal on Scientific Computing}, 18:1--22, 1997.

\bibitem{Psakis2011}
Georgios Psakis, Jo~Mailliet, Christina Lang, Lotte Teufel, Lars-Oliver Essen,
  and Jon Hughes.

\end{thebibliography}
%\addcontentsline{toc}{chapter}{References}
\end{document}